\documentclass{aa}  
\usepackage{graphicx}
\usepackage{amsmath,amssymb}
\usepackage{txfonts}
\begin{document}

   \title{LABOCA mapping of the infrared dark cloud MSXDC G304.74+01.32
   \thanks{This publication is based on data acquired with the Atacama 
Pathfinder Experiment (APEX). APEX is a collaboration between the 
Max-Planck-Institut f\"ur Radioastronomie, the European Southern Observatory,
and the Onsala Space Observatory.}}

   \author{O. Miettinen \and J. Harju}

   \offprints{O. Miettinen}

   \institute{Department of Physics, P.O. Box 64, 00014 University of 
Helsinki, Finland\\
              \email{oskari.miettinen@helsinki.fi}}

\date{Received ; accepted}

\authorrunning{Miettinen \& Harju}
\titlerunning{LABOCA mapping of the IRDC G304.74}

  \abstract
   {Infrared dark clouds (IRDCs) likely represent very early stages
of high-mass star/star cluster formation.}
   {The aim is to determine the physical properties and spatial
distribution of dense clumps in the IRDC MSXDC G304.74+01.32 (G304.74), and 
bring these characteristics into relation to theories concerning the origin of
IRDCs and their fragmentation into clumps and star-forming cores.} 
   {G304.74 was mapped in the 870 $\mu$m dust continuum with the 
LABOCA bolometer on APEX. The 870 $\mu$m map was compared with the 1.2 mm 
continuum map of the cloud by B\'eltran et al. (2006). Archival MSX and
IRAS infrared data were used to study the nature and properties of the 
submillimetre clumps within the cloud. The H$_2$ column densities were 
estimated using both the 870 $\mu$m dust emission and the MSX 8 $\mu$m 
extinction data. The obtained values were compared with near-infrared
extinction which could be estimated along a few lines of sight. 
We compared the clump masses and their spatial distribution in G304.74 
with those in several other recently studied IRDCs.}
   {Twelve clumps were identified from the 870 $\mu$m dust continuum map. 
Three of them are associated with the MSX and IRAS point sources. 
Moreover, one of the clumps (SMM 6) is associated with two MSX 8 $\mu$m 
point-like sources. Thus, there are 8 clumps within G304.74 which are 
not associated with mid-infrared (MIR) emission. The H$_2$ column 
densities derived from the dust continuum and extinction data are similar. 
The comparison suggests that the dust temperature may be elevated (20-30 K) 
near the southern end of the cloud, whereas the starless clumps in the centre 
and in the north are cool ($T_{\rm d}\sim15$ K). 
There is a high likelihood that the clump mass distributions in 
G304.74 and in several other IRDCs represent the samples of the same parent
distribution. In most cases the spatial distributions of clumps in 
IRDCs do not deviate significantly from random distributions.} 
   {G304.74 contains several massive clumps that are not
associated with MIR emission. On statistical grounds it is likely that
some of them are or harbour high-mass starless cores (HMSCs). The fact that 
the clump mass distributions (resembling the high-mass stellar IMF), 
and in some cases also the random-like spatial distributions, seem to be 
comparable between different IRDCs, is consistent with the idea that the 
origin of IRDCs, and their further sub-fragmentation down to scales of clumps 
is caused by supersonic turbulence in accordance with results from giant 
molecular clouds.}

    \keywords{Stars: formation - ISM: clouds - ISM: structure - 
Radio continuum: ISM - Submillimetre}

    \maketitle
%

\section{Introduction}

\subsection{Infrared dark clouds}

The so-called infrared dark clouds (IRDCs) were discovered by the Infrared 
Space Observatory (ISO; \cite{perault1996}) and the Midcourse Space Experiment
(MSX; \cite{egan1998}); the clouds were detected as dark absorption objects 
against the bright mid-infrared (MIR) Galactic background radiation. 
Based on the MSX 8 $\mu$m data of the Galactic plane from $l=0-360\degr$ 
and $\left| b \right|\leq5\degr$, Simon et al. (2006a) identified almost 
11\,000 IRDC candidates. Recently, Peretto \& Fuller (2009) used Spitzer 
satellite data to extract about 9000 new IRDC candidates.

Molecular line and dust continuum studies of IRDCs have shown that they are
cold ($T<25$ K), dense ($n({\rm H_2})\gtrsim10^5$ cm$^{-3}$, 
$N({\rm H_2})\gtrsim10^{22}$ cm$^{-2}$), and massive 
($\sim10^2-10^5$ M$_{\sun}$) structures with sizes of $\sim1-15$ pc 
(e.g., \cite{carey1998}; \cite{simon2006b}; \cite{rathborne2006}, hereafter 
RJS06; \cite{du2008}; \cite{vasyunina2009}). Most IRDCs are filamentary 
(e.g., \cite{peretto2009}), and contain density enhancements, or 
clumps\footnote{We prefer to use the term ``clump'' according to e.g., Ragan
et al. (2009, hereafter RBG09), rather than the term ``core'' which was used
by e.g., RJS06. Clumps have masses and sizes (radii) of $\sim10-10^3$ 
M$_{\sun}$ and $\sim0.1-1$ pc, respectively (e.g., \cite{bergin2007}).}, 
that are visible in (sub)millimetre dust continuum maps 
(e.g., \cite{carey2000}; \cite{garay2004}; \cite{ormel2005}; RJS06). 
The cold clumps (i.e., clumps unassociated with MSX 8 $\mu$m emission) 
identified by RJS06 have typical sizes and masses of $\sim0.5$ pc and 
$\sim120$ M$_{\sun}$, respectively. Because IRDCs have clumpy structures, 
they are likely to be in an early stage of fragmentation 
(e.g., RBG09 and references therein).

The radial galactocentric distribution of IRDCs peaks at $R_{\rm GC}=5$ kpc in
the 1st Galactic quadrant, and at $R_{\rm GC}=6$ kpc in the 4th quadrant, 
which correspond to the location of the Scutum-Centaurus spiral arm 
(see \cite{simon2006b} and \cite{jackson2008}). 
This, together with the fact that IRDCs have sizes
and masses similar to those of warm massive cluster-forming regions 
(e.g., \cite{lada2003}; \cite{motte2003}), has led to the suggestion
that IRDCs represent the very early stages of high-mass star/star cluster 
formation. Indeed, several studies have found signs of ongoing star formation 
within IRDCs. These include CH$_3$OH and H$_2$O masers (\cite{beuther2002b}; 
\cite{pillai2006a}; \cite{wang2006}; \cite{ellingsen2006}), outflow signatures 
(\cite{beuther2005}; \cite{beuthersrid2007}; \cite{sakai2008}), 
and associated infrared sources (e.g., \cite{rathborne2005}; 
\cite{beutherstein2007}; \cite{chambers2009}; RBG09). 
Cold, dense clumps are suggested to host/represent precursors of
{\sl hot} molecular cores, i.e., high-mass starless cores (HMSCs; e.g.,
\cite{sridharan2005}; \cite{beuther2007}). A few hot cores have been found
within IRDCs (\cite{rathborne2007}, 2008), but it is important to increase the
sample in order to study the sequence of events. It should be noted that the
formation of hot cores may not be universal in IRDCs - some of them seem to
form only low- to intermediate-mass stars (e.g., \cite{vanderwiel2008}).  

\subsection{Infrared dark cloud MSXDC G304.74+1.32}

The IRDC studied in this paper was designated as MSXDC G304.74+1.32 
(hereafter, G304.74) by Simon et al. (2006a). 
G304.74 was observed by Beltr{\'a}n et al. (2006) with the SIMBA bolometer 
array on SEST at 1.2 mm. Beltr{\'a}n et al. (2006) 
identified eight millimetre clumps within the cloud, out of which 
four were found to be associated with MSX point sources. 
The distance to the cloud, 2.4 kpc, is a kinematic distance estimated
from the velocity of the CS line observed by Fontani et al. (2005).
The galactocentric distance is $\sim7.4$ kpc.

G304.74 was chosen for the present study because its relatively close distance
allows for reasonably good spatial resolution in order to study its 
substructure. Moreover, a relatively high number of clumps (8) were already 
identified from the cloud (see above), and thus it was considered an 
appropriate object for studying IRDC fragmentation.

\vspace{0.7cm}

In this paper, we present the results of our 870 $\mu$m dust continuum 
mapping of G304.74. The paper pays special attention to the clumpy
structure of the cloud, and the clump mass distribution, and thereby addresses
the fragmentation of IRDCs. The observations and data-reduction procedures are
described in Sect. 2. The observational results are presented in Sect. 3. 
The MSX 8 $\mu$m optical thicknesses toward the submm peaks are 
derived in Sect. 4. In Sect. 5, we describe the methods used to derive the 
physical properties of the observed clumps. In Sect. 6, we discuss the results
of our study, and in Sect. 7, we summarise our main conclusions.

\section{Observations and data reduction}

\subsection{Submillimetre dust continuum}

The 870 $\mu$m dust continuum observations toward G304.74 were 
carried out on 25--27 April 2009 with the 295 channel bolometer array LABOCA 
(Large APEX Bolometer Camera) on APEX.
The central frequency of the instrument is 345 GHz, and the bandwidth is 
60 GHz. The half-power beam width (HPBW) of the telescope is $18\farcs6$ 
($\sim0.22$ pc at 2.4 kpc) at the frequency used. 
The total field of view (FoV) of LABOCA is $11\farcm4$. 
The instrument and its observing modes are described in Siringo et al. (2009).

Absolute flux calibration was achieved through observations of the 
planets Mars, Uranus, and Neptune (the primary calibrators for LABOCA), 
and the star CW Leo as secondary calibrator. The uncertainty due to flux 
calibration was estimated to be $\sim10\%$.
The telescope focus and pointing were checked using the
planet Saturn, the star $\eta$ Carinae, and the H$_2$O maser source 
B13134 (305.80-00.24).
The submm zenith opacity was determined using the sky-dip method and
the values varied from 0.20 to 0.38, with a median value of 0.30. The 
observations were thus conducted in fair weather conditions. 

The observations were made using the the on-the-fly (OTF) mapping mode with
a scanning speed of $3\arcmin$ s$^{-1}$. The size of the OTF map is 
$34\arcmin \times 34\arcmin$. This was achieved by 204 subscans of $34\arcmin$
in length (parallel with the R.A. axis), spaced by $10\arcsec$. The area was
mapped five times, with a total observing time of 5.2 h. In this manner, 
a uniform rms noise level of $\sigma=0.03$ Jy beam$^{-1}$ was reached in the
central $\sim15\arcmin \times 15\arcmin$ area which covers the target source.

The data reduction was performed using the BoA (Bolometer Array Analysis
Software) software package according to guidelines in the BoA User and
Reference 
Manual (2007)\footnote{{\tt http://www.astro.uni-bonn.de/boawiki/Boa}}.
The data reduction included flat-fielding, flagging bad/dark channels and data
according to telescope speed and acceleration, correcting for the atmospheric
opacity, division into subscans, baseline subtractions and median-noise
removal (correction for sky noise), despiking, and filtering-out of 
the low frequencies of the $1/f$-noise. Finally, the five individual maps were
coadded. 

We note that even the source model was iteratively used in the 
reduction process, the resulting final map (Fig.~\ref{figure:LABOCA}, left 
panel) has negative artefacts (``holes'') around regions of bright emission. 
The depths of these negative holes are $\sim15-50\%$ ($\sim30\%$ on average) of
the nearest peak object brightness, and thus they are likely to introduce 
additional uncertainty in the source flux densities. We estimate that the 
total flux density uncertainty due to calibration and negative artefacts is 
in the range $\sim20-50\%$. These uncertainties are not, however, taken into 
account in the analyses presented in this paper.

\subsection{Complementary near-, mid-, and far-infrared data}

The target source ($l=304\fdg74$, $b=1\fdg32$) is not included in the GLIMPSE
(3.6, 4.5, 5.8, 8.0 $\mu$m) and MIPSGAL (24 and 70 $\mu$m) surveys of Spitzer
which cover the Galactic latitudes $\left| b \right|\leq1\degr$. 
We have used the near-infrared 2MASS ($J$, $H$, and $K_{\rm s}$) survey data
archive (\cite{skrutskie2006}), data products from the MSX survey
(\cite{price2001}; \cite{egan2003}), and the IRAS (Infrared Astronomical 
Satellite) MIR and far-infrared (FIR) observations. 

\section{Observational results}

\subsection{Clump identification}

The obtained LABOCA map is presented 
in the left panel of Fig.~\ref{figure:LABOCA}.
The cloud has a filamentary appearance and extends over about 13\farcm2, i.e., 
about 9.2 pc at the cloud's distance. 
In order to identify clumps in the LABOCA map, we employed the two-dimensional
\texttt{clumpfind} algorithm, \texttt{clfind2d}, developed by Williams et al. 
(1994). The \texttt{clfind2d} routine determines the peak position, the 
FWHM (full width at half maximum) size (not corrected for beam size), 
and the peak and total integrated flux density of clump based on specified 
contour levels. The algorithm requires two input parameters: 1)
the intensity threshold, i.e., the lowest contour level, which determines the 
minimum emission to be included into clumps; and 2) the stepsize which 
determines the required ``contrast'' between two clumps to be considered as 
different objects. We set both the intensity threshold and the stepsize to 
$2\sigma$ of the background noise level. With these settings, 
G304.74 divides into 12 clumps. 

We note that the total number and size of clumps identified by 
\texttt{clumpfind} is quite sensitive to the selected contour levels.
The selected $2\sigma$ contour levels turned out to give the  
best agreement with the identification by eye. This selection is also
recommended by Williams et al. (1994). See also Pineda et al. (2009) 
for a recent discussion of clump identification with \texttt{clumpfind}.
Pineda et al. (2009) concluded that small changes in the threshold and/or 
stepsize values can lead to important changes in the number of identified 
clumps.

\subsection{Observed properties of the clumps}

\begin{figure*}
\centering
\includegraphics[width=10cm]{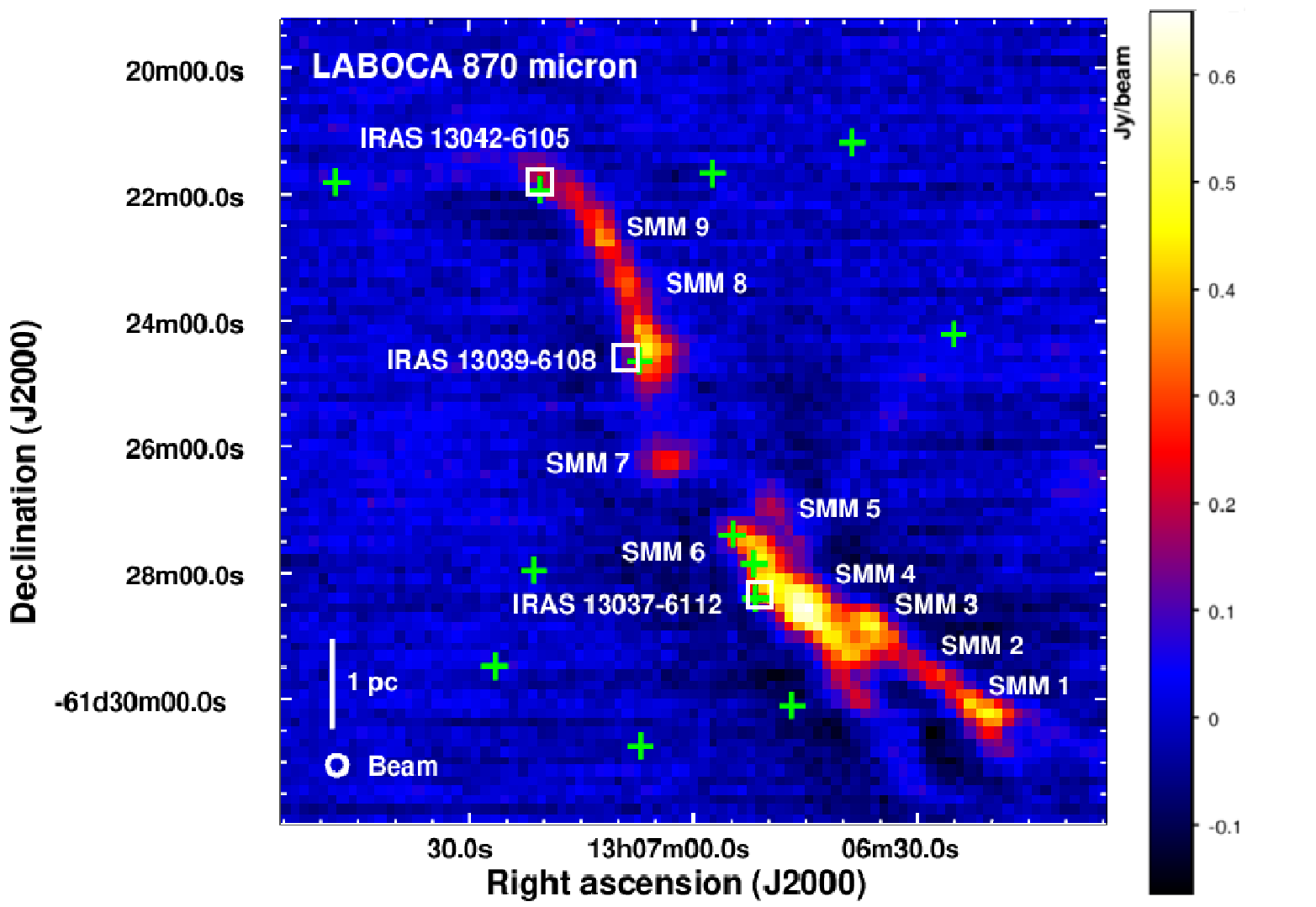}
\includegraphics[width=8.3cm]{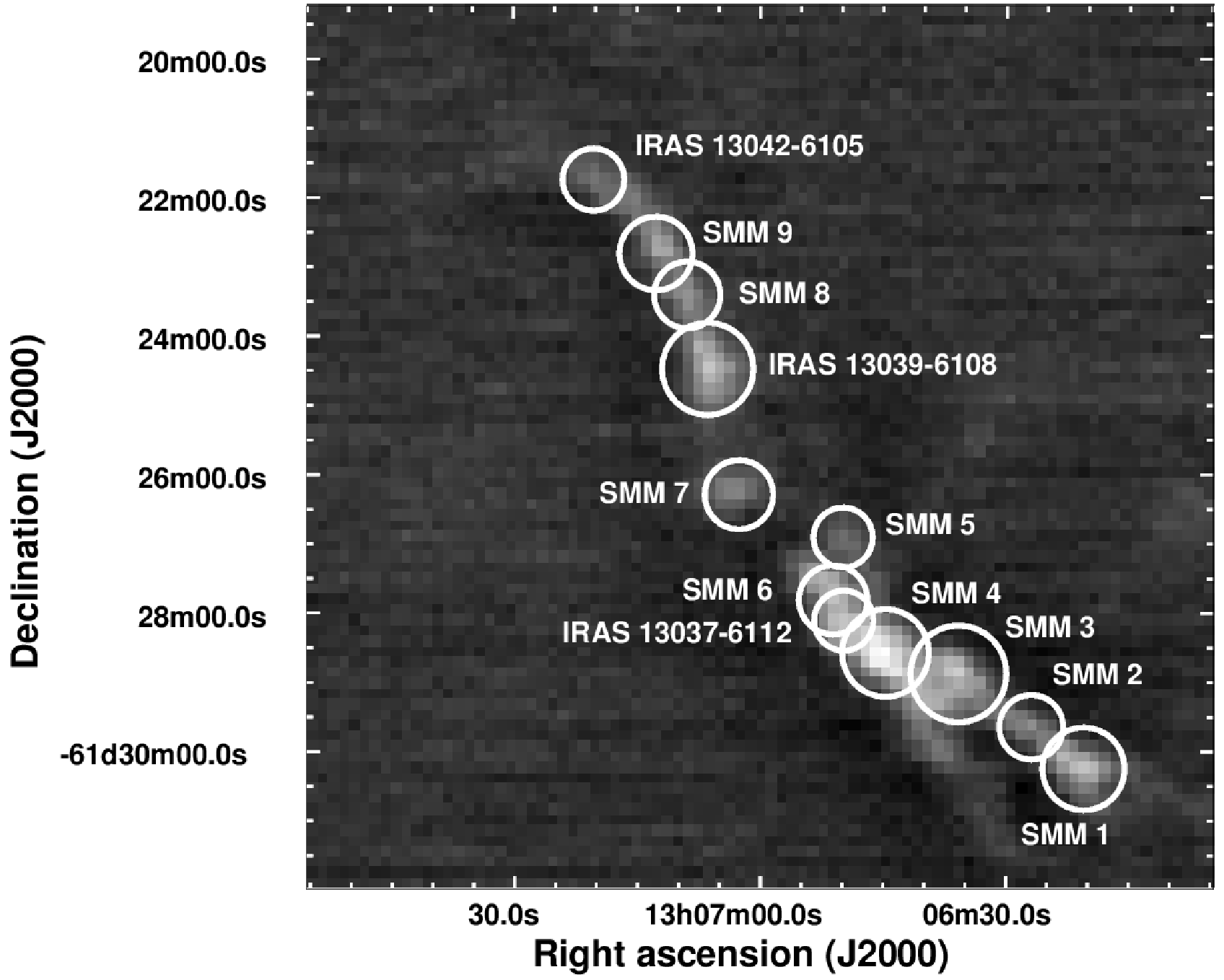}
\caption{\textbf{Left}: LABOCA map of the 870 $\mu$m dust continuum 
emission from the IRDC G304.74. The green plus signs and white boxes mark the 
positions of the MSX 8 $\mu$m sources in the field (cf.~Fig.~\ref{figure:MSX}) 
and IRAS point sources, respectively. The intensity range is from -0.17 to 
0.66 Jy beam$^{-1}$, as indicated in the colour bar. 
The beam HPBW ($18\farcs6$) and scale-bar are shown in the bottom left.
\textbf{Right:} Grey-scale image of 870 $\mu$m emission shown in the 
left panel with the locations and sizes of the submm clumps determined by the
\texttt{clumpfind} algorithm. The circles represent the area associated with 
each clump but not the shape.}
\label{figure:LABOCA}
\end{figure*}

All the clumps have a peak flux density $>5\sigma$ (i.e., $>0.15$ Jy 
beam$^{-1}$) relative to the local background. The coordinates, peak and
integrated flux densities, and deconvolved angular FWHM diameters 
($\theta_{\rm s}$) are listed in Cols.~(2)-(6) of Table~\ref{table:clumps}. 
The values of $\theta_{\rm s}$ correspond to the beam corrected FWHM size.
Column~(7) lists the effective radius, $R_{\rm eff}$. 
This radius is defined as $R_{\rm eff}=\sqrt{A/\pi}$, where $A$ is
the projected area with the clump boundaries. The corresponding 
circles are overlaid on the LABOCA map in the right panel of 
Fig.~\ref{figure:LABOCA}. They are centred on the dust peak positions 
(Cols.~(2) and (3)). Note that the \texttt{clumpfind} 
algorithm makes no assumption about the shape of the clump.

\begin{figure*}
\centering
\includegraphics[width=9cm]{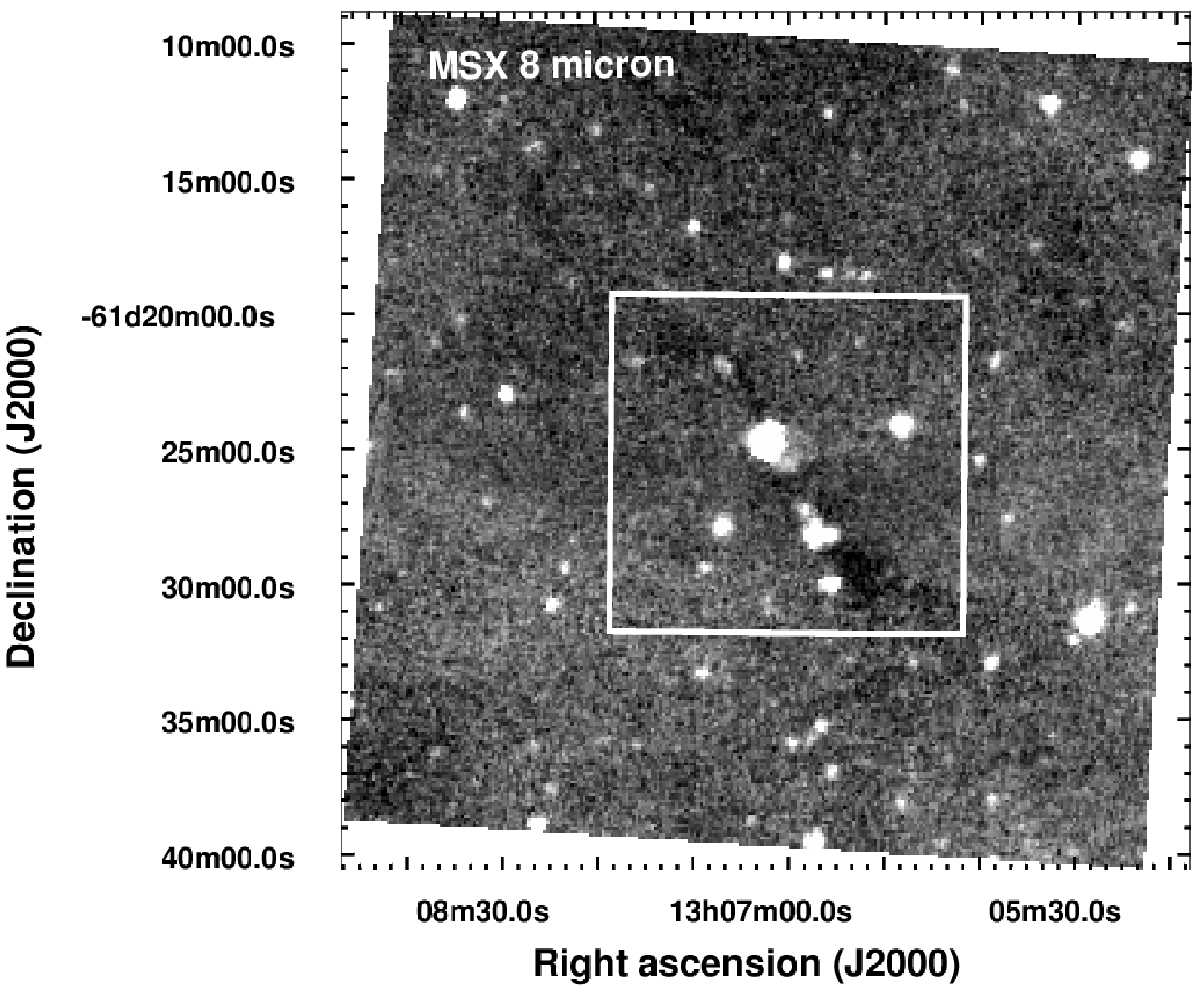}
\includegraphics[width=9cm]{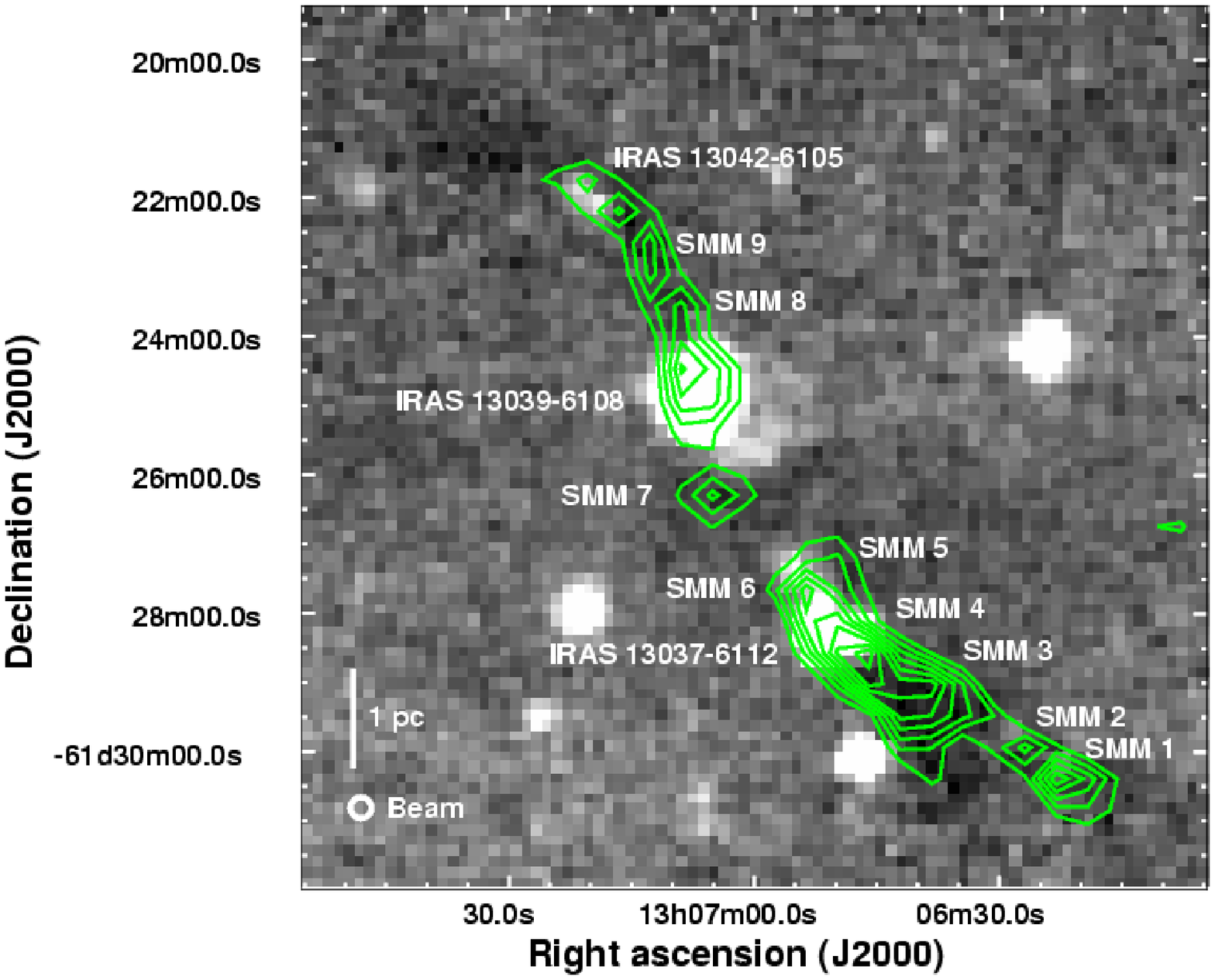}
\caption{\textbf{Left:} Wide-field ($0\fdg5 \times 0\fdg5$, i.e., 
$\sim21\times21$ pc at 2.4 kpc) MSX 8 $\mu$m image showing the MIR extinction 
of the IRDC G304.74. \textbf{Right:} Blow-up of the left panel showing the 
same region as in Fig.~\ref{figure:LABOCA}. 
The MSX 8 $\mu$m image is overlaid with contours of LABOCA 870 $\mu$m 
emission. The intensity range of a grey-scale image is from 
$1.86\times10^{-7}$ to $2.32\times10^{-5}$ W~m$^{-2}$~sr$^{-1}$. 
The contours are 0.06 ($2\sigma$) to 0.66 Jy beam$^{-1}$, in steps of 0.06 
Jy beam$^{-1}$. The beam HPBW ($18\farcs3$) and scale-bar are shown in the
bottom left.}
\label{figure:MSX}
\end{figure*}

The left panel of Fig.~\ref{figure:MSX} shows the wide-field MSX 8 $\mu$m 
image (extracted from the MSX Galactic Plane Survey 
images\footnote{{\tt http://irsa.ipac.ipac.caltech.edu}}) towards G304.74. 
The MSX 8 $\mu$m image zooming to the dark cloud with 
LABOCA contours is shown in the right panel of Fig.~\ref{figure:MSX}.
There is a good agreement between the morphologies of
the 870 $\mu$m continuum emission and the 8 $\mu$m extinction visible as a
dark lane in the MSX image. The resolution of the MSX image, $18\farcs3$, 
is similar to that of the 870 $\mu$m LABOCA map.

Three of the submm clumps are associated with sources both from 
the MSX and IRAS point source catalogues (within $\sim6-23\arcsec$ 
and $\sim4-20\arcsec$ from the dust peak position, respectively; 
see Fig.~\ref{figure:LABOCA}, left). 
These are designated according to the IRAS names. The remaining nine clumps 
are named SMM 1, SMM 2, etc. One of the submm clumps, SMM 6, appears to be 
associated with two 8 $\mu$m sources from the MSX catalogue. 
The stronger 8 $\mu$m source is located at $\sim6\arcsec$ from the 
dust peak position (0.46 Jy beam$^{-1}$), whereas the weaker one is located at 
$\sim19\arcsec$ from the closest dust emission peak (0.41 Jy beam$^{-1}$) 
within SMM 6. Furthermore, the eastern edge of SMM 4 is bright
in 8 $\mu$m as the MSX source associated with IRAS 13037-6112 extends about 
55\arcsec west. Thus there are seven clumps which are completely dark in the
MIR.

The flux densities at MIR and FIR wavelengths retrieved from the 
MSX (8.28, 12.13, 14.65, 21.34 $\mu$m) and IRAS (12, 25, 60, 100 $\mu$m) 
archives for all the IRAS sources and for the two 8 $\mu$m point sources 
associated with SMM 6 are listed in Tables~\ref{table:MSX} and 
~\ref{table:IRAS}, respectively.

\begin{table*}
\caption{Submillimetre clumps identified by the \texttt{clumpfind} algorithm 
in the IRDC G304.74.}
\begin{minipage}{2\columnwidth}
\centering
\renewcommand{\footnoterule}{}
\label{table:clumps}
\begin{tabular}{c c c c c c c c}
\hline\hline 
 & \multicolumn{2}{c}{Peak position} & $S_{870}^{\rm peak}$\footnote{The peak and total flux densities have estimated uncertainties of $\sim20-50\%$, depending on the intensity of negative holes around the clump (see Sect.~2.1).} & $S_{870}^a$ & $\theta_{\rm s}$ & $R_{\rm eff}$\footnote{The effective radius, $R_{\rm eff}=\sqrt{A/\pi}$, where $A$ is the area assigned to the clump, i.e., area contained within a $2\sigma$ contour. The radius has not been deconvolved from the beam size.} & 1.2 mm clump No.\footnote{Corresponding 1.2 mm clump in Beltr\'an et al. (2006).}\\
Name & $\alpha_{2000.0}$ [h:m:s] & $\delta_{2000.0}$ [$\degr$:$\arcmin$:$\arcsec$] & [Jy beam$^{-1}$] & [Jy] & [\arcsec] & [\arcsec] & (Beltr\'an et al. 2006)\\
\hline
SMM 1 & 13 06 20.5 & -61 30 15 & 0.46 & 2.0 & 36 & 36 &\\
SMM 2 & 13 06 26.9 & -61 29 39 & 0.30 & 1.0 & 38 & 28 &\\
SMM 3 & 13 06 35.8 & -61 28 53 & 0.50 & 3.4 & 45 & 42 & 1\\
SMM 4 & 13 06 44.7 & -61 28 35 & 0.66 & 4.1 & 38 & 38 & 3\\
IRAS 13037-6112 & 13 06 49.8 & -61 28 07 & 0.47 & 1.6 & 24 & 26 & 2\\
SMM 5 & 13 06 49.9 & -61 26 55 & 0.19 & 0.7 & 28 & 26&\\
SMM 6 & 13 06 51.1 & -61 27 49 & 0.46 & 1.8 & 27 & 30 & 4\\
SMM 7 & 13 07 02.6 & -61 26 18 & 0.26 & 0.9 & 24 & 30 & 8\\
IRAS 13039-6108 & 13 07 06.4 & -61 24 29 & 0.47 & 2.6 & 37 & 40 & 5\\
SMM 8 & 13 07 08.9 & -61 23 25 & 0.31 & 1.1 & 26 & 29 & \\
SMM 9 & 13 07 12.7 & -61 22 49 & 0.38 & 1.6 & 35 & 32 & 6\\
IRAS 13042-6105 & 13 07 20.3 & -61 21 45 & 0.21 & 0.8 & 28 & 27 & 7\\
\hline 
\end{tabular} 
\end{minipage}
\end{table*}

\begin{table*}
\caption{MSX point sources in the IRDC G304.74.}
\begin{minipage}{2\columnwidth}
\centering
\renewcommand{\footnoterule}{}
\label{table:MSX}
\begin{tabular}{c c c c c c}
\hline\hline 
 & & $S_{8.28}$ & $S_{12.13}$ & $S_{14.65}$ & $S_{21.34}$\\
Associated clump & MSX6C designation & [Jy] & [Jy] & [Jy] & [Jy]\\
\hline
IRAS 13037-6112 & G304.7728+01.3431 & $0.876\pm0.037$ & $<0.876\pm0.076$ & $1.027\pm0.076$ & $<1.969\pm0.142$\\
SMM 6 & G304.7738+01.3522 & $0.213\pm0.011$ & - & $<0.753\pm0.061$ & $2.972\pm0.193$\\
SMM 6 & G304.7800+01.3597 & $0.184\pm0.010$ & - & - & -\\
IRAS 13039-6108 & G304.8074+01.4037 & $1.875\pm0.077$ &  $2.125\pm0.123$ & $<0.730\pm0.063$ & $<1.678\pm0.129$\\
IRAS 13042-6105 & G304.8366+01.4472 & $0.147\pm0.010$ & - & - & -\\
\hline 
\end{tabular} 
\end{minipage}
\end{table*}

\begin{table}
\caption{IRAS point sources in the IRDC G304.74.}
\centering
\renewcommand{\footnoterule}{}
\label{table:IRAS}
\begin{tabular}{c c c c c}
\hline\hline 
 & $S_{12}$ & $S_{25}$ & $S_{60}$ & $S_{100}$\\
Name &  [Jy] & [Jy] & [Jy] & [Jy]\\
\hline
IRAS 13037-6112 & 1.84 & 7.41 & 64.96 & $<196.5$\\
IRAS 13039-6108 & 3.99 & 7.43 & 105.6 & 196.5\\
IRAS 13042-6105 & $<1.31$ & 0.67 & $<4.97$ & $<196.5$\\
\hline 
\end{tabular} 
\end{table}

\section{IRDC extinction and the 8 $\mu$m optical thicknesses}

\subsection{Principle}

The observed 8 $\mu$m intensity towards an IRDC, $I_{\rm IRDC}^{\rm obs}$, 
can be written as (e.g., \cite{bacmann2000}) 

\begin{equation}
\label{eq:intensity}
I_{\rm IRDC}^{\rm obs}=I_{\rm bg}e^{-\tau_{\rm 8 \mu m}}+I_{\rm fore} \,,
\end{equation}
where $I_{\rm bg}$ is the 8 $\mu$m intensity of the background emission, 
$\tau_{\rm 8 \mu m}$ is the 8 $\mu$m optical thickness,
and $I_{\rm fore}$ is the 8 $\mu$m intensity contribution from foreground 
material. Equation~(\ref{eq:intensity}) can be inverted to get the value of 
$\tau_{\rm 8 \mu m}$ as 

\begin{equation}
\label{eq:tau}
\tau_{\rm 8 \mu m}=-\ln \left(\frac{I_{\rm IRDC}^{\rm obs}-I_{\rm fore}}{I_{\rm bg}}\right) \,.
\end{equation}

\begin{figure}[!h]
\resizebox{\hsize}{!}{\includegraphics[angle=270]{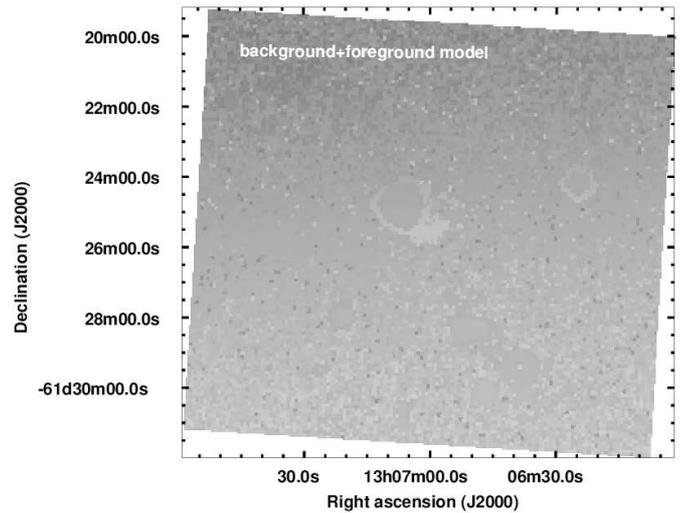}}
\caption{Median filter estimate of $I_{\rm MIR}$. The intensity scale is linear
from $1.20\times10^{-6}$ to $1.26\times10^{-6}$ W~m$^{-2}$~sr$^{-1}$, with both
mean and median being $1.23\pm0.02\times10^{-6}$ W~m$^{-2}$~sr$^{-1}$.}
\label{figure:MIR}
\end{figure}

Following the notation of Bacmann et al. (2000) and Peretto \& Fuller (2009), 
the observed MIR intensity around the IRDC is 

\begin{equation}
\label{eq:MIR}
I_{\rm MIR}=I_{\rm bg}+I_{\rm fore} \,.
\end{equation}
To obtain an estimate for the value of $I_{\rm MIR}$, we used the technique
of spatial median filtering (\cite{simon2006a}). In this method, 
$I_{\rm MIR}$ for each pixel is estimated by taking the median
of intensities inside a surrounding filter region.
The size of the filter must be larger than the size of IRDC; the method will
only capture emission fluctuations on scales larger than the size of
the cloud. Given the size of G304.74 ($\gtrsim13\arcmin$, i.e., over 130 
6\arcsec pixels of the MSX map), we used a circular spatial filter with a
radius of $15\arcmin$ (the same as used by Simon et al. (2006a)). 
We note that the total size of the image used in the filtering process 
was $0\fdg5\times0\fdg5$ (see Fig.~\ref{figure:MSX}, left). 
Figure~\ref{figure:MIR} shows the obtained model of $I_{\rm MIR}$. 
Both the mean and median value of $I_{\rm MIR}$ are $1.23\pm0.02\times10^{-6}$ 
W~m$^{-2}$~sr$^{-1}$, where the uncertainty is the standard deviation of 
the pixel values. The above value was adopted in the following analysis.

\subsection{Background and foreground estimation}

To estimate the contribution of $I_{\rm bg}$ and $I_{\rm fore}$ to the 
value of $I_{\rm MIR}$ (Eq.~(\ref{eq:MIR}), and 
Fig.~\ref{figure:MIR}), we used the similar analysis as used by Butler \& Tan 
(2009, their section 3.1). We assumed that the radial distribution of
Galactic 8 $\mu$m PAH (polycyclic aromatic hydrocarbon) emission has the same 
exponential form as the radial distribution of molecular gas surface density
in the Galaxy (the radial scale length is $H_R=3.5\pm1.0$ kpc; 
\cite{williams1997}). We note that this distribution is similar to that of 
the Galactic surface density of OB associations (\cite{mckee1997}), 
which was used by Butler \& Tan (2009, their Eq.~(3)). The assumption 
that the radial distribution of PAHs follows the molecular gas is supported
by the results of Mattila et al. (1999) who found that the UIR (unidentified
infrared) emission band intensities in the disk of the galaxy NGC 891 closely
follow those of the CO emission and 1.3 mm dust continuum emission. However, 
UIR band intensities were found to be quite different from that of the 
H$\alpha$ emission (see Fig.~5 of Mattila et al. (1999)). 
Moreover, Kahanp{\"a\"a} et al. (2003) found strong correlation between the 
Galactic UIR bands and CO emission.
To estimate the vertical distribution of PAH emission, we used the Galactic
vertical distribution of CO emission and adopted the scale length 
$z_{1/2}=70\pm10$ pc (\cite{dame1987}; \cite{bronfman1988}; 
\cite{malhotra1994}).

\begin{figure}[!h]
\resizebox{\hsize}{!}{\includegraphics{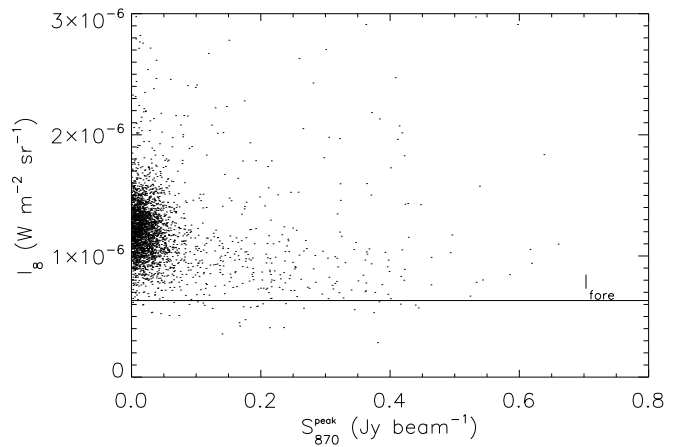}}
\caption{MSX 8 $\mu$m versus LABOCA 870 $\mu$m intensity for G304.74. 
The horizontal line marks the 8 $\mu$m foreground intensity, 
$I_{\rm fore}=6.32\times10^{-7}$ W~m$^{-2}$~sr$^{-1}$, derived from the 
foreground intensity ratio, $f_{\rm fore}$ (see main text).}
\label{figure:msxandlaboca}
\end{figure}

We first calculated the so-called ``foreground intensity ratio'', 
$f_{\rm fore}$, which is defined as the ratio of the column of PAH emission
between the observer and the cloud and the total column through the Galaxy
along the same line of sight, up to the galactocentric distance 16 kpc.
For G304.74, we obtained $f_{\rm fore}=0.514\pm0.05$, where the uncertainty 
represents the mimimum-maximum error based on the uncertainties in the scale 
lengths (see above). The distance of the Sun from
the Galactic Centre was assumed to be 8.5 kpc in this calculation. The 
intensity of the background emission is then given by

\begin{equation}
\label{eq:background}
I_{\rm bg}=(1-f_{\rm fore})I_{\rm MIR} \,.
\end{equation}
The resulting values are 
$I_{\rm bg}=5.98\pm0.62\times10^{-7}$ W~m$^{-2}$~sr$^{-1}$ and 
$I_{\rm fore}=6.32\pm0.65\times10^{-7}$ W~m$^{-2}$~sr$^{-1}$.
The errors in $I_{\rm bg}$ and $I_{\rm fore}$ are propagated from 
the standard deviation of $I_{\rm MIR}$ and the uncertainty in $f_{\rm fore}$.
The above value of $I_{\rm fore}$ is in good agreement with the lowest 
observed 8 $\mu$m intensities as demonstrated in  
Fig.~\ref{figure:msxandlaboca}. We note that the anti-correlation between 
the submm and MIR intensities allows for a determination of $I_{\rm bg}$ 
and $I_{\rm fore}$ (e.g., \cite{johnstone2003}, Fig.~3 therein).
However, the correlation shown in Fig.~\ref{figure:msxandlaboca} is not very
clear and thus only the $f_{\rm fore}$-method is considered in the present 
paper (see Sect.~6.1 for further discussion).

The observed 8 $\mu$m intensities toward the submm dust continuum 
peak positions, and the 8 $\mu$m optical thicknesses calculated from
Eq.~(\ref{eq:tau}), are listed in Cols.~(2) and (3) of 
Table~\ref{table:extinction}. The formal error in 
$\tau_{\rm 8 \mu m}^{\rm peak}$ was calculated by propagating the errors
in $I_{\rm fore}$ and $I_{\rm bg}$. In order to determine 
$I_{\rm IRDC}^{\rm obs}$ values in the submm peak positions, the MSX image
was gridded with $9\farcs3$ pixels, i.e., the LABOCA pixel size. We also 
estimated the possibility that part of the observed 8 $\mu$m intensity  
originates from the bright surroundings of a MIR dark clump. 
The measured point spread function (PSF) in the MSX image is a
Gaussian with a FWHM of $22\arcsec$. Convolution with this PSF reduces the
breadths of dark filaments, and, for the narrowest of them, increases
the minimum intensity in the middle. We estimate that for a
completely opaque source with a diameter of $\lesssim30\arcsec$, the MIR
radiation ``leakage'' from the surroundings amounts to $\gtrsim20\%$ of the
background level. This contribution was, however, neglected in the above
analysis which uses only the peak (minimum) 8 $\mu$m intensities. 
Also, most of the MIR dark clumps are larger than $30\arcsec$ 
(see Table~\ref{table:clumps}, Cols.~(6) and (7)).

\subsection{Dust temperature estimates}

Because dust continuum emission is optically thin at (sub)mm 
wavelengths, the 870 $\mu$m radiation intensity is given by

\begin{equation}
\label{eq:flux}
I_{870}\simeq B_{870}(T_{\rm d})\tau_{870}\,,
\end{equation}
where $B_{870}(T_{\rm d})$ is the Planck function with dust 
temperature $T_{\rm d}$. According to the Ossenkopf \& Henning (1994, 
hereafter OH94) dust model used in the present paper (see Sect.~5.1 for more 
details), the ratio of dust opacities per unit dust mass at 
8.8\footnote{Note, for the MSX band A filter the 50\% cutoffs are at 6.8 and 
10.8 $\mu$m (central wavelength is 8.8 $\mu$m), whereas the isophotal central 
wavelength of the filter is 8.28 $\mu$m.} and 870 $\mu$m, 
$\kappa_8/\kappa_{870}$, and thus the correponding ratio of the optical 
thicknesses, $\tau_8/\tau_{870}$, is about 865. 
The obtained ratio is consistent with the results of 
Johnstone et al. (2003) and Ormel et al. (2005), who found that 
(assuming $T_{\rm d}=15$ K) $\kappa_{8}/\kappa_{850}\sim640$ and 870, 
respectively. By using this ratio, and the $\tau_8$ values derived above,
estimates for the dust temperatures, $T_{\rm d}$, towards the submm peaks
can be derived from Eq.(~\ref{eq:flux}). The resulting values are listed
in Col.~(4) or Table~\ref{table:extinction}. The quoted errors are the 
minimum-maximum errors derived from the uncertainty in $\tau_8$.

The uncertainties of the $T_{\rm d}$ estimates are very large in the 
southern part (SMM 1, SMM 2; in the case of SMM 3 and SMM 4 no reasonable 
estimate could be obtained because of the large error in $\tau_8$). 
On the other hand, the estimates with moderate formal errors in the central and
northern part of the cloud suggest low temperatures of slightly above
10 K. It should be noted, however, that there are large uncertainties
concerning the dust opacities and the determination of the $\tau_8$. This
issue will be raised again in Sect.~6.2.

In view of these uncertainties, dust temperatures
estimated above should be compared with previous temperature
determinations from molecular lines. Observations towards several
other IRDC clumps (H$_2$CO, \cite{carey1998}; CH$_3$CCH, \cite{teyssier2002}; 
NH$_3$, \cite{pillai2006b}; \cite{sakai2008}) provide gas kinetic
temperatures of $T_{\rm kin}\approx10-20$ K. The assumption that 
$T_{\rm d}=T_{\rm kin}$ is probably valid in dense clouds 
($n({\rm H_2})\gtrsim10^5$ cm$^{-3}$; e.g., \cite{goldsmith1978}). 
Based on these considerations, we assume in what
follows that in general $T_{\rm d}=15$ K. Exceptions are made in the cases of
two IRAS sources, for which temperatures can be derived from the
spectral energy distributions (SEDs). We will also separately consider
the possibility that $T_{\rm d}$ is elevated in the four southern clumps 
SMM 1--4.

\begin{table}
\caption{The observed 8 $\mu$m intensity, the peak optical 
thickness, and the dust temperature toward the submm peak positions.}
\begin{minipage}{1\columnwidth}
\centering
\renewcommand{\footnoterule}{}
\label{table:extinction}
\begin{tabular}{c c c c}
\hline\hline 
   & $I_{\rm IRDC}^{\rm obs}$ & $\tau_{\rm 8 \mu m}^{\rm peak}$ & $T_{\rm d}$\\
Name & [$10^{-7}$ W~m$^{-2}$~sr$^{-1}$] & & [K]\\
\hline
SMM 1 & 9.89 & $0.52\pm0.21$ & $30.6\pm17.0$\\
SMM 2 & 10.39 & $0.38\pm0.19$ & $28.0\pm19.6$\\
SMM 3 & 11.73 & 0.10\footnote{No error given because it is greater than the
actual value.} & -\footnote{The value of $T_{\rm d}$ could not be 
reasonably estimated.}\\
SMM 4 & 10.97 & $0.25\pm0.17$ & -$^b$\\
SMM 5 & 8.03 & $1.25\pm0.40$ & $10.1\pm2.8$\\
SMM 7 & 8.08 & $1.22\pm0.39$ & $12.0\pm3.7$\\
SMM 8 & 7.58 & $1.56\pm0.53$ & $11.5\pm3.7$\\
SMM 9 & 6.99 & $2.19\pm0.98$ & $10.8\pm4.3$\\
\hline 
\end{tabular} 
\end{minipage}
\end{table}

\section{Physical parameters of the clumps}

\subsection{Spectral energy distributions of the IRAS sources}

\begin{figure}[!h]
\resizebox{\hsize}{!}{\includegraphics{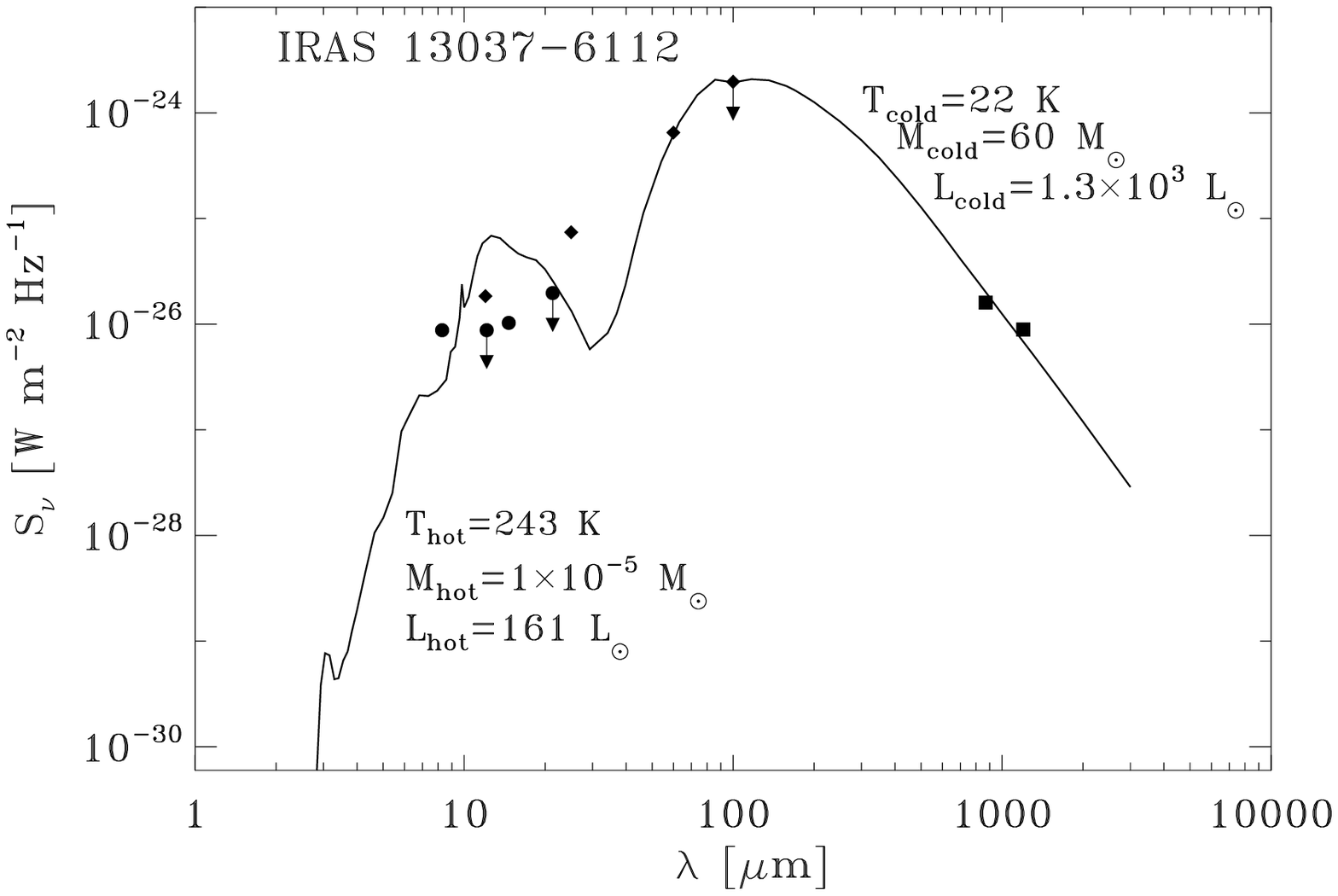}}
\resizebox{\hsize}{!}{\includegraphics{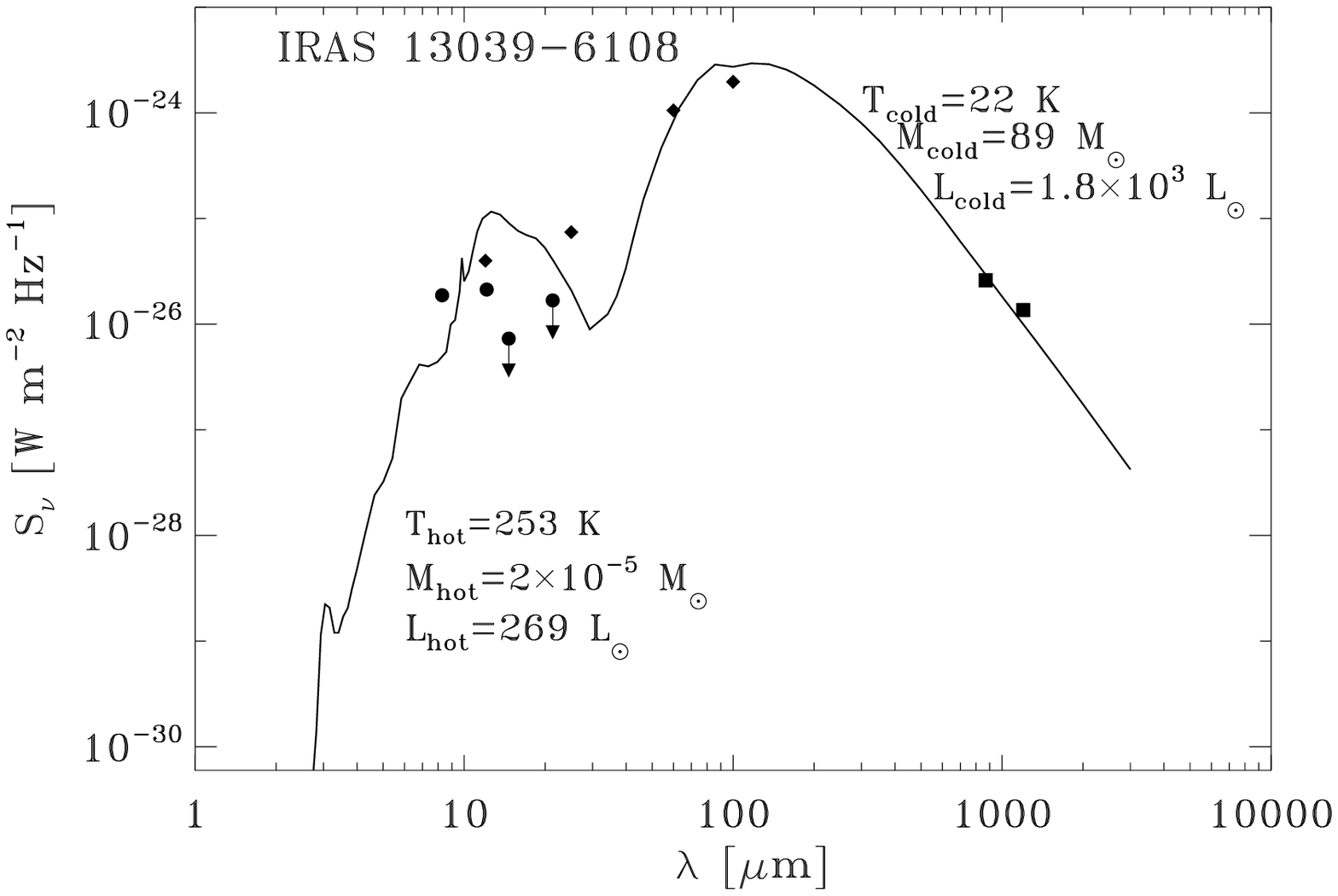}}
\caption{Best-fit SEDs of IRAS 13037-6112 and IRAS 13039-6108.
MSX data points are shown by circles, IRAS data points are shown by diamonds, 
and LABOCA and SIMBA (sub)mm data points are indicated by squares. 
Upper limits are indicated by arrows. The solid lines in both plots 
represent the sum of two (cold$+$hot) components (see Cols.~(4) and (5) of 
Table~\ref{table:sed}). The absorption features appearing at $\sim3.1$ 
and 10 $\mu$m are attributed to the H$_2$O ice and silicates, respectively 
(OH94 and references therein).}
\label{figure:sed}
\end{figure}

The MSX, IRAS, LABOCA, and SIMBA data were used to fit the spectral energy 
distributions (SEDs) of IRAS 13037-6112 and IRAS 13039-6108.
The SIMBA 1.2 mm flux densities of these sources are 0.89 and 1.36 Jy,
respectively (\cite{beltran2006}; Table~2 therein). 
Note that there is not enough data points for IRAS 13042-6105 
in order to construct a reasonable SED (e.g., most of its IRAS flux densities
are only upper limits, see Table~\ref{table:IRAS}). 
The derived SEDs are shown in Fig.~\ref{figure:sed}. 
The least-squares fitting routine used in the derivation of the 
SED\footnote{The SED fitting routine was originally written by 
J.~Steinacker.} minimises the sum 
$\sum_{i=1}^N \left[\log_{10}\left(S_{\nu}^{\rm obs}(\lambda_i)\right)-\log_{10}\left(S_{\nu}^{\rm mod}(\lambda_i)\right)\right]^2$, where $N$ is the number 
of data points (7 and 8 for IRAS 13037-6112 and 13039-6108, respectively), 
$S_{\nu}^{\rm obs}(\lambda_i)$ is the observed flux density, and 
$S_{\nu}^{\rm mod}(\lambda_i)$ is the model flux density.
In both cases, the data were fitted by a 
two-temperature composite model. It was assumed that both components at 
different temperatures emit as a blackbody modified by the 
wavelength-dependent dust opacity, $\kappa_{\lambda}$ (see below). 
The best-fit model SEDs overestimate
the flux densities at $\sim12-20$ $\mu$m, but underestimate them at $\sim8$
and 25 $\mu$m. It should be noted that the flux densities included in 
the SEDs are measured using telescopes with different beam sizes. Thus the 
flux densities obtained for extended sources are not fully comparable, and 
this can in part explain discrepancies between MSX ($18\farcs3$) 
and IRAS ($\sim2\arcmin$ at 12 $\mu$m to $\sim4\arcmin$ at 100 $\mu$m) fluxes 
at 12 and $\sim20-25$ $\mu$m. On the other hand, SIMBA and LABOCA 
flux densities at $\lambda=1.2$ mm and 0.87 mm refer to clump areas 
($R_{\rm eff}$ is typically $\sim30\arcsec$) derived by \texttt{clumpfind}. 
$R_{\rm eff}$ values are similar for both IRAS 13037-6122 and IRAS 13039-6108 
($\sim30-40\arcsec$ or $\sim 0.35-0.47$ pc). Assuming that 
the emission in the IRAS bands is confined in the region of the submm
clump, the characteristic spatial scale associated with the SEDs is 
$\lesssim 0.5$ pc.

We have adopted a dust-to-gas mass ratio of $R_{\rm d}=1/100$, a value 
which has often been used in the IRDC studies (e.g., RJS06; 
\cite{vasyunina2009}; \cite{parsons2009}). However,   
this value can differ from 1/100. For instance, Draine et al. (2007)
determined a value of $R_{\rm d}\approx1/186$ based on observed depletions
of heavy elements in the Galaxy. Dust opacities we have adopted correspond
to a MRN size distribution with thick ice 
mantles\footnote{In cold, dense interiors of the
IRDCs, dust grains are supposed to be coagulated and covered by icy mantles 
(e.g., \cite{peretto2009} and references therein). 
This is supported by e.g., the observed depletion of H$_2$CO  
(\cite{carey1998}, 2000), CO (\cite{pillai2007}; 
\cite{zhang2009}), and CS (\cite{beuther2009}) in IRDCs. 
Moreover, Butler \& Tan (2009) found some evidence for dust opacity changes 
within IRDCs, which could be caused by ice mantle formation and grain growth.} 
at a gas density of $n_{\rm H}=10^5$ cm$^{-3}$ (OH94).
The resulting SED parameters are given in Table~\ref{table:sed}.
The total (cold+hot) mass and the integrated bolometric luminosity are given
in Cols.~(2) and (3) of Table~\ref{table:sed}, respectively.
The temperatures of the two components are listed
in Cols.~(4) and (5). In Cols.~(6) and (7), we give the mass and 
luminosity fractions of the cold component versus the total mass and 
luminosity. Column~(8) lists the mass-to-luminosity ratio, 
$M_{\rm tot}/L_{\rm bol}$, which is an evolutionary indicator of the clump as
it is expected to decrease with time. The envelope mass decreases during 
the star formation process, and the luminosity of the embedded star or stellar
cluster rises (e.g., \cite{sridharan2002}). We note that the adopted 
dust opacity model (corresponding to particles coated with thick ice mantles) 
is not likely to be appropriate for hot dust, and therefore the total 
bolometric luminosity for the hot component should be taken with caution. 
The MIR spectral features (such as PAH emission) also cause the fit to the hot
component being more uncertain than the fit to the cold part of
the spectrum. However, this should not alter the fact that for both IRAS 
sources the mass of the hot component is negligible 
($\sim1-2\times10^{-5}$ M$_{\sun}$), and thus the bulk of the 
material is cold ($M_{\rm cold}/M_{\rm tot}\sim1$). The bolometric 
temperature of IRAS 13039-6108 ($T_{\rm cold}\simeq22$ K) is in good 
agreement with the rotation temperature of 18 K derived from C$^{17}$O by 
Fontani et al. (2005).

\begin{table*}
\caption{Results of the SED fits.}
\begin{minipage}{2\columnwidth}
\centering
\renewcommand{\footnoterule}{}
\label{table:sed}
\begin{tabular}{c c c c c c c c c}
\hline\hline 
    & $M_{\rm tot}$ & $L_{\rm bol}$ & $T_{\rm cold}$ & $T_{\rm hot}$ &  &  & $M_{\rm tot}/L_{\rm bol}$\\ 
Source & [M$_{\sun}$] & [$10^3$ L$_{\sun}$] & [K] & [K] & $M_{\rm cold}/M_{\rm tot}$ & $L_{\rm cold}/L_{\rm bol}$ & [M$_{\sun}$/L$_{\sun}$]\\
\hline
IRAS 13037-6112 & 60 & 1.5 & 22 & 243 & $\sim1$ & 0.90 & 0.04\\
IRAS 13039-6108 & 89 & 2.1 & 22 & 253 & $\sim1$ & 0.87 & 0.04\\
\hline 
\end{tabular} 
\end{minipage}
\end{table*}

\subsection{Linear sizes, masses, and H$_2$ number densities}

The clump linear radii were computed from the effective radii 
listed in Col.~(7) of Table~\ref{table:clumps}.
The clump masses (gas+dust mass, $M_{\rm cont}$) were calculated
from their integrated 870 $\mu$m flux density, $S_{870}$ 
(Table~\ref{table:clumps}, Col.~(5)), assuming that the thermal dust emission
is optically thin (\cite{hildebrand1983}):

\begin{equation}
\label{eq:mass}
M_{\rm cont}=\frac{S_{870}d^2}{B_{870}(T_{\rm d})\kappa_{870}R_{\rm d}} \,,
\end{equation}
where $d$ is the source distance. For IRAS 13037-6112 and 13039-6108,
we assumed the dust temperatures to be the same as their bolometric 
temperatures, $T_{\rm cold}\simeq22$ K,
resulting from SED fits (see Table~\ref{table:sed}, Col.~(4)). 
For all the other clumps (SMM 1, SMM 2,\ldots, and IRAS 13042-6105),
it was assumed that $T_{\rm d}=15$ K (Sect.~4.3). 
We assumed that $\kappa_{870}\simeq0.17$ m$^2$ kg$^{-1}$. This value is 
interpolated from OH94, see Sect. 5.1. As in the SED 
fits, the value $1/100$ is adopted for $R_{\rm d}$.

The volume-averaged H$_2$ number densities, $\langle n({\rm H_2}) \rangle$,
were calculated assuming a spherical geometry for the clumps, using the formula

\begin{equation}
\label{eq:density}
\langle n({\rm H_2}) \rangle=\frac{\langle \rho \rangle}{\mu_{{\rm H_2}}m_{\rm H}}\,,
\end{equation}
where $\langle \rho \rangle=M_{\rm cont}/\left(4/3 \pi R_{\rm eff}^3\right)$ is the matter density, $\mu_{\rm H_2}=2.8$ is the mean molecular weight 
per H$_2$ molecule (assuming a 10\% helium abundance), and $m_{\rm H}$ is the 
mass of the hydrogen atom.
The obtained radii, masses, and volume-averaged H$_2$ number densities are
given in Cols.~(2), (3), and \textbf{(7)} of Table~\ref{table:parameters}, 
respectively. The typical density in the clumps is likely to be higher
than the volume-averaged value because of substructure.
Note that our masses are in general larger than those derived by 
Beltr{\'a}n et al. (2006) who assumed a dust temperature of 30 K and a 
$\kappa_{\rm 1.2mm}$ of 0.1 m$^2$ kg$^{-1}$ in their mass estimates. 
They also obtained higher densities because of using the FWHM 
radii (instead of effective radii) and a mean molecular weight of 2.29.

\begin{table*}
\caption{Linear radii, masses, and H$_2$ column and volume-averaged number
densities of the submm clumps within the IRDC G304.74.}
\begin{minipage}{2\columnwidth}
\centering
\renewcommand{\footnoterule}{}
\label{table:parameters}
\begin{tabular}{c c c c c c c c}
\hline\hline 
       & $R_{\rm eff}$ & $M_{\rm cont}$ & $N_{870}({\rm H_2})$ & $N_{8}({\rm H_2})$ &  & $ \langle n({\rm H_2}) \rangle $\\ 
Source & [pc] & [M$_{\sun}$] & [$10^{22}$ cm$^{-2}$] & [$10^{22}$ cm$^{-2}$] & $N_{870}/N_{8}$ & [$10^4$ cm$^{-3}$]\\  
\hline
SMM 1\footnote{By assuming $T_{\rm d}=30$ K (see Sect.~4.3 and Table~\ref{table:extinction}), $M_{\rm cont}=39$ M$_{\sun}$, $N_{870}({\rm H_2})=0.75\pm0.05\times10^{22}$ cm$^{-2}$, $N_{870}/N_{8}=1.0\pm0.1$, and $\langle n({\rm H_2}) \rangle=0.2\times10^4$ cm$^{-3}$.} & 0.42 & 107 & $2.06\pm0.13$ & $0.74\pm0.30$ & $2.8\pm1.1$ & 0.7\\
SMM 2\footnote{By assuming $T_{\rm d}=30$ K, 
$M_{\rm cont}=20$ M$_{\sun}$, $N_{870}({\rm H_2})=0.49\pm0.05\times10^{22}$ 
cm$^{-2}$, $N_{870}/N_{8}=0.9\pm0.1$, and $\langle n({\rm H_2}) \rangle=0.3\times10^4$ cm$^{-3}$.} & 0.33 & 53 & $1.35\pm0.14$ & $0.54\pm0.27$ & $2.5\pm1.3$ & 0.7\\
SMM 3\footnote{By assuming $T_{\rm d}=30$ K, 
$M_{\rm cont}=66$ M$_{\sun}$, $N_{870}({\rm H_2})=0.82\pm0.05\times10^{22}$ 
cm$^{-2}$, $N_{870}/N_{8}=5.9\pm0.4$, and $\langle n({\rm H_2}) \rangle=0.3\times10^4$ cm$^{-3}$.} & 0.49 & 182 & $2.24\pm0.13$ & 0.14\footnote{No error given because it is greater than the actual value.} & $16\pm0.9$ & 0.7\\
SMM 4\footnote{By assuming $T_{\rm d}=30$ K, 
$M_{\rm cont}=80$ M$_{\sun}$, $N_{870}({\rm H_2})=1.08\pm0.05\times10^{22}$ 
cm$^{-2}$, $N_{870}/N_{8}=3.0\pm2.0$, and $\langle n({\rm H_2}) \rangle=0.4\times10^4$ cm$^{-3}$.} & 0.44 & 219 & $2.96\pm0.13$ & $0.36\pm0.24$ & $8.2\pm5.5$ & 1.2\\
IRAS 13037-6112 & 0.30 & 48 & $1.17\pm0.07$ & - & - & 0.8\\
SMM 5 & 0.30 & 37 & $0.85\pm0.13$ & $1.79\pm0.57$ & $0.5\pm0.2$ & 0.6\\
SMM 6 & 0.35 & 96 & $2.06\pm0.13$ & - & - & 1.0\\
SMM 7 & 0.35 & 48 & $1.17\pm0.14$ & $1.74\pm0.56$ & $0.7\pm0.2$ & 0.5\\
IRAS 13039-6108 & 0.47 & 78 & $1.18\pm0.08$ & - & - & 0.3\\
SMM 8 & 0.34 & 59 & $1.39\pm0.13$ & $2.23\pm0.76$ & $0.6\pm0.2$ & 0.7\\
SMM 9 & 0.37 & 85 & $1.70\pm0.13$ & $3.13\pm1.40$ & $0.5\pm0.2$ & 0.8\\
IRAS 13042-6105 & 0.31 & 43 & $0.94\pm0.13$ & - & - & 0.7\\
\hline 
\end{tabular} 
\end{minipage}
\end{table*}

\subsection{H$_2$ column densities}

The H$_2$ column densities, $N({\rm H_2})$, towards the submm peaks 
were estimated using 1) the dust emission at 870 $\mu$m, and 2) the dust 
extinction at 8 $\mu$m. In the first method, the LABOCA intensities were 
converted to $N({\rm H_2})$ using the equation

\begin{equation}
\label{eq:N_H2}
N_{870}({\rm H_2})=\frac{I_{870}^{\rm dust}}{B_{870}(T_{\rm d})\mu_{\rm H_2} m_{\rm H}\kappa_{870}R_{\rm d}} \,,
\end{equation}
where $I_{870}^{\rm dust}=S_{870}^{\rm peak}/\Omega_{\rm beam}$ is the observed
dust peak surface brightness ($\Omega_{\rm beam}$ is the beam solid angle),
which is related to the peak flux density via 1 Jy/18\farcs6 beam
$=1.085\times10^{-18}$ W m$^{-2}$ Hz$^{-1}$ sr$^{-1}$. 
The dust temperature values used were identical to those 
adopted in the mass estimates (Eq.~(\ref{eq:mass})). The uncertainty in 
$N_{870}({\rm H_2})$ was calculated by propagating the uncertainty in 
$I_{870}^{\rm dust}$, and thus reflects only the $1\sigma$ rms noise in the
870 $\mu$m map. 

The MSX 8 $\mu$m optical thicknesses (Table~\ref{table:extinction}, 
Col.~(3)) were used to estimate the peak H$_2$ column densities by applying
the formula 

\begin{equation}
\label{eq:N_H2_MSX}
N_{8}({\rm H_2})=\frac{\tau_{\rm 8 \mu m}^{\rm peak}}{\sigma_{\lambda}}\,,
\end{equation}
where $\sigma_{\lambda}$ is the dust extinction cross-section per H$_2$ 
molecule. According to the OH94 dust model we have used 
(see Sect. 5.1), the value $\sigma_{\lambda}=\kappa_{\lambda}\mu_{\rm H_2}m_{\rm H}R_{\rm d}\approx7.0\times10^{-23}$ cm$^2$ at 8.8 $\mu$m. 
The uncertainty in $N_{8}({\rm H_2})$ was calculated by 
propagating the uncertainty in $\tau_{\rm 8 \mu m}^{\rm peak}$ and do not 
include the systematic error resulting from the estimate of $\sigma_{\lambda}$.
The obtained H$_2$ column densities, and the $N_{870}/N_{8}$ ratios, 
are given in Cols.~(4)-(6) of Table~\ref{table:parameters}. 

\subsection{Extinction estimates with 2MASS}

\begin{figure} 
\includegraphics[height=9cm, width=14.5cm]{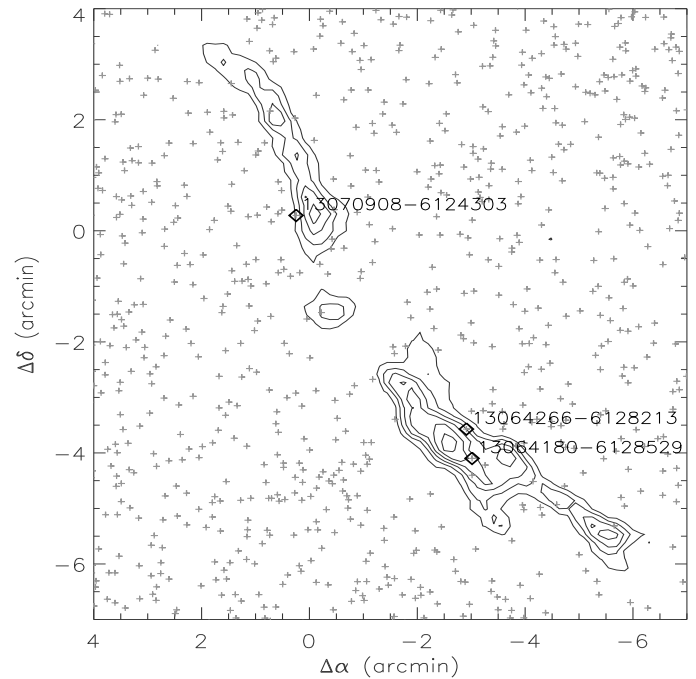} 
\caption{LABOCA dust continuum map (contours) with 2MASS point sources
(plus signs). The heavily reddened 2MASS stars within the cloud 
boundaries are marked with diamonds. The contour levels go from 0.1 to 0.6
by 0.1 Jy beam$^{-1}$.}
\label{figure:laboca+2mass} 
\end{figure}

We examined if extinctions estimated from the $JHK_{\rm s}$ photometry of
2MASS stars lying in the region can be used to calibrate the H$_2$ column
densities derived above\footnote{The 2MASS All-Sky Point Source
Catalog (PSC) used here have been made available at {\tt
http://irsa.ipac.caltech.edu/}}. A rarefaction of 2MASS point sources 
can be discerned in the cloud region and its immediate vicinity, and it is not
possible to derive an extinction map of such detail that a comparison between 
the LABOCA map would be meaningful. Altogether 14 2MASS point sources with 
good photometric quality in all three bands lie within the LABOCA contour 0.1 
Jy beam$^{-1}$. Three of them show $J-H$ and $H-K_{\rm s}$ colours 
characteristic of giant stars reddened by substantial columns of interstellar 
dust. This judgement is based on their locations near the standard reddening 
line on the $J-H$ vs. $H-K_{\rm s}$ plot, and on the fact that their 
$J-K_{\rm s}$ excesses, $E_{J-K}$, are larger than 0.6 (see below).
The three stars are the 2MASS PSC objects 13064180-6128529,
13064266-6128213, and 13070908-6124303 (see Table~\ref{table:2mass_stars} and
Fig.~\ref{figure:laboca+2mass}). The first
two lie near the clump SMM  4 and are likely to represent background K or M
giants. The third is found very close ($\sim10\arcsec$) to IRAS 13039-6108
and is somewhat too bright for a class III giant in view of the distance and
extinction. Because its NIR colours ($J-H=2.56\pm0.03$, 
$H-K_{\rm s}=1.31\pm0.03$) are consistent with an YSO (young stellar object) 
candidate with NIR excess (\cite{matsuyanagi2006}), the source is possibly 
a NIR counterpart of IRAS 13039-6108.

\begin{table*}
\caption{Reddened 2MASS point sources within the IRDC G304.74.}
\begin{minipage}{2\columnwidth}
\centering
\renewcommand{\footnoterule}{}
\label{table:2mass_stars}
\begin{tabular}{ccccccccc} 
\hline\hline 
& $J$ & $H$ & $K_{\rm s}$ & $J-K_{\rm s}$ & $E_{J-K}$\footnote{Assuming $(J-K)_0=0.63-1.13$ corresponding to spectral classes K0 III-M3 III.} & $N_{\rm 2MASS}({\rm H_2})$\footnote{Assuming $N({\rm H_2})/E_{J-K}=5.4\times10^{21}\, {\rm cm^{-2}}\, {\rm mag}^{-1}$.} & $N_{870}({\rm H_2})$ & \\
2MASS designation & [mag] & [mag] & [mag] & [mag] & [mag] & [$10^{22}$ cm$^{-2}$] & [$10^{22}$ cm$^{-2}$] & $N_{870}/N_{\rm 2MASS}$ \\ 
\hline  
13064180-6128529 & $14.15\pm0.04$ & $12.79\pm0.05$ & $12.04\pm0.04$ & $2.11\pm0.06$ &  $1.0-1.5$  & $0.53-0.80$ & 2.3 & $2.9-4.3$\\             
13064266-6128213 & $15.98\pm0.08$ & $14.42\pm0.04$ & $13.53\pm0.05$ & $2.44\pm0.10$ &  $1.3-1.8$  & $0.71-0.98$ & 1.4 & $1.4-2$\\            
13070908-6124303 & $14.26\pm0.02$ & $11.70\pm0.02$ & $10.39\pm0.02$ & $3.87\pm0.03$ &  $2.7-3.2$  & $1.5-1.7$ & 0.62 & $\sim0.4$\\                
\hline 
\end{tabular}
\end{minipage}
\end{table*}

The H$_2$ column density ranges given in Col.~(7) of
Table~\ref{table:2mass_stars} are derived using intrinsic colours 
for K0 III--M3 III giants (\cite{bessell1988}), and the relationship 
$N({\rm H_2})/E_{J-K}=5.4\times10^{21}\, {\rm cm^{-2}}\, {\rm mag}^{-1}$
(\cite{bohlin1978}; \cite{mathis1990}; \cite{harjunpaa1996}). The colour
excesses, $E_{J-K}$, measure the total amount of dust in front of the
stars, so they are affected by dust in the foreground, and possibly
also in the background of the cloud. Therefore, the $N({\rm H_2})$
values are upper limits for the contribution of the cloud itself.  A
nearby line of sight ($l=304\fdg75$, $b=1\fdg25$) has been included in the
investigation of 3-D distribution of the extinction by Marshall et
al. (2006). Their results suggest a foreground
extinction of $A_K \sim 0.4$ ($E_{J-K}\sim0.6$; \cite{marshall2006};
\cite{mathis1990}) up to the cloud's distance 2.4 kpc.
 
The LABOCA 870 $\mu$m intensities in the directions of 13064180-6128529,
13064266-6128213, and 13070908-6124303 are 0.49, 0.31, and $0.13\pm
0.03$ Jy beam$^{-1}$, respectively. The conversion to the H$_2$ column 
densities using Eq.~(\ref{eq:N_H2}) with $T_{\rm d}=15$ K gives 
$2.3\times10^{22}$, $1.4\times10^{22}$, and $0.62\times10^{22}$ cm$^{-2}$.
The first two values are not consistent with the upper limits derived from
$E_{J-K}$ (see Table~\ref{table:2mass_stars}, Col.~(9)).
In the case of 13064180-6128529, the H$_2$ column density derived from LABOCA
870 $\mu$m data can be brought down to conform with 2MASS estimate by 
increasing the dust temperature to $T_{\rm d}=30$ K, whereas for the location
of 13064266-6128213 an increase to $T_{\rm d}=20$ K would be sufficient. 
The latter value is in good agreement with the bolometric temperature of 
IRAS 13037-6112 ($T_{\rm cold}\simeq22$ K, see Table~\ref{table:sed}, 
Col.~(4)) which lies (in the plane of the sky) quite close to 13064266-6128213.
The 8 $\mu$m intensities in the directions of 13064180-6128529 and 
13064266-6128213 are $1.03\times10^{-6}$ and $1.22\times10^{-6}$ 
W~m$^{-2}$~sr$^{-1}$, respectively. These correspond to 8 $\mu$m 
optical thicknesses of $0.41\pm0.19$ and 0.02 (here the associated 
error is larger than the value), respectively (see Eq.~(\ref{eq:tau})). 
Using Eq.~(\ref{eq:N_H2_MSX}), the corresponding H$_2$ column densities 
become $0.59\pm0.27\times10^{22}$ and $0.03\times10^{22}$ cm$^{-2}$, 
respectively. The first value is consistent with the 2MASS estimate, 
whereas the latter value is much lower. 

The above results clearly show that it is important to know the dust 
temperature in order to accurately determine the H$_2$ column density from the
submm dust continuum emission (Eq.~(\ref{eq:N_H2})). 
On the other hand, the MIR absorption and 2MASS extinction 
methods require several uncertain assumptions, such as the dust model and the
corresponding dust extinction cross-section, contribution of the foreground
emission, and the relation between $N({\rm H_2})$ and $E_{J-K}$. Moreover, 
as the 2MASS extinction could be estimated only along a few lines of sight, 
the present statistics is very poor. Thus, the data presented here do not 
offer a firm conclusion about the most reliable method to determine the 
H$_2$ column densities. We note that the empirical
relationship between $N_{\rm H}$ and $A_J$ derived by Vuong et al. 
(2003) in $\rho$ Oph, used recently by Marshall et al. (2009), implies 
$N({\rm H_2})/E_{J-K}\simeq4.7\times10^{21}$ cm$^{-2}\, {\rm mag}^{-1}$. 
This would make the H$_2$  column densities about 13\% smaller than those 
estimated above.

\section{Discussion}

\subsection{Estimating the background and foreground MIR emission}

The average MIR emission around the cloud, 
$I_{\rm MIR}=1.23\pm0.02\times10^{-6}$ W~m$^{-2}$~sr$^{-1}$,
was estimated in Sect.~4 by using median filtering. Vasyunina et al. (2009)
estimated $I_{\rm MIR}$ in the vicinity of several IRDCs
directly from the 8 $\mu$m images. They pointed out that it is difficult to
control the process of median filtering if large ($\gtrsim10\arcmin$) filters
are needed. On the other hand, if the filter is too small (comparable to the
cloud size), the value of $I_{\rm MIR}$ will be underestimated because 
it will be affected by the cloud itself (\cite{butler2009}).
Because the largest diameter of G304.74 is $\sim 13\arcmin$, we tried
the direct method by choosing manually MIR emission patches (free of strong
MIR emission sources) in the close vicinity of the IRDC. We used three 
different patches with angular sizes of $6\farcs7 \times 3\farcs3$, 
$4\farcs8 \times 2\farcs3$, and $5\farcs7 \times 1\farcs8$.
The mean and standard deviation of the 8 $\mu$m intensity within these regions
were $1.17\pm0.18\times10^{-6}$, $1.21\pm0.18\times10^{-6}$, and 
$1.25\pm0.21\times10^{-6}$ W~m$^{-2}$~sr$^{-1}$, respectively.
The mean and standard deviation of these three are 
$1.21\pm0.11\times10^{-6}$ W~m$^{-2}$~sr$^{-1}$. This result is in excellent
agreement with the value obtained from median filtering.

Based on the observed anti-correlation between the 850 $\mu$m and 8 $\mu$m 
intensities (cf. Fig.~\ref{figure:msxandlaboca}), Johnstone et al. (2003) and 
Ormel et al. (2005) found that $I_{\rm fore}\simeq I_{\rm bg}$ in the case of
IRDC G11.11-0.12 and the W51 IRDC, respectively. 
Peretto \& Fuller (2009) constrained $I_{\rm fore}$ by the
requirement that both MIR absorption and 1.2 mm emission should
give the same $N({\rm H_2})$. In this method, it is assumed that the true
8 $\mu$m opacity can be calculated from the millimetre emission 
(see Eq.~(3) of Peretto \& Fuller (2009)). Then, the value of $I_{\rm fore}$
can be calculated in terms of $I_{\rm MIR}$ (see our Eq.~(\ref{eq:tau})). 
Peretto \& Fuller (2009) also showed that on the average the background
emission is approximately equal to the foreground emission (their Eq.~(5)). 
The drawback in this approach is the uncertainty in the ratio of the mass 
absorption coefficients at MIR and (sub)mm wavelengths 
($R_{\kappa}$ in Eq.~(3) of Peretto \& Fuller (2009)). In the present study, 
we estimated the contributions of background and foreground emission 
using the foreground intensity ratio (Sect.~4.2). We also ended up 
with the result that $I_{\rm fore}\simeq I_{\rm bg}$.
Butler \& Tan (2009) and RBG09 estimated that 
$I_{\rm fore}\simeq(0.1-0.5)\times I_{\rm bg}$ ($f_{\rm fore}$-method) and 
$I_{\rm fore}=(2-5)\times I_{\rm bg}$ (850 $\mu$m--8 $\mu$m anti-correlation)
for their samples of IRDCs, respectively.

\subsection{Comparison of H$_2$ column densities determined from dust 
continuum emission and extinction data}

The H$_2$ column densities derived from submm emission and
MIR absorption are mostly in good agreement (within a factor of $\sim2$; 
Table~\ref{table:parameters}, Cols.~(4) and (5)). 
This suggests that the dust parameters used in 
Eqs.~(\ref{eq:N_H2}) and (\ref{eq:N_H2_MSX}) are reasonable.
The dust temperature, $T_{\rm d}$, is likely to show spatial variations, and
this causes uncertainties to column density estimates based on
submm emission (Eq.~(\ref{eq:N_H2})), but does not affect the 8 $\mu$m 
absorption method (Eq.~(\ref{eq:N_H2_MSX})). The greatest differences between
the two $N({\rm H_2})$ values are found toward positions with the highest 
$N_{870}({\rm H_2})$ values. These are SMM 1, 3, and 4, where the 
$N_{870}({\rm H_2})/N_{8}({\rm H_2})$ ratios are $2.8\pm1.1$, $16\pm0.9$, 
and $8.2\pm5.5$, respectively (Col.~(6) of 
Table~\ref{table:parameters}). Some of these discrepancies could (in part)
be remedied, e.g., by increasing the dust temperature (cf.~footnote in 
Table~\ref{table:parameters}). It seems possible that
the dust temperature is higher than 15 K in the four clumps (SMM 1--4)
near southwestern tip of the cloud. On the other hand, 
Vasyunina et al. (2009) showed that the extinction method becomes unreliable 
at very high column densities, but this should happen only at 
$N({\rm H_2})\gtrsim10^{23}$ cm$^{-2}$.   

However, there are also uncertainties in deriving $N_{8}({\rm H_2})$,
in particular related to the contamination by the foreground emission (here
done by using the foreground intensity ratio, see Sect.~4), and the dust
extinction cross-section (reliable only within a factor of $\sim2$, e.g.,
\cite{ragan2006}). Also, the value of $\kappa_{\lambda}$, needed in the 
calculation of $N_{870}({\rm H_2})$ (Eq.~(\ref{eq:N_H2})) has an uncertainty
similar to that of $\sigma_{\lambda}$. The dust-to-gas mass ratio could also 
differ from the adopted value 1/100 as mentioned in Sect.~5.1. 
Thus, the direct comparison of $N({\rm H_2})$ derived from dust continuum and
extinction data should be taken with caution. 
We note that Parsons et al. (2009) found for their
sample of IRDCs that $N_{8}({\rm H_2})$ (MSX) and $N_{850}({\rm H_2})$ (SCUBA)
agree within an order of magnitude. The moderate correspondence may be partly
explained by the omission of the foreground emission.

\subsection{Nature of submm clumps}

The MSX 8 $\mu$m emission associated with molecular clumps suggests 
the presence of protostars. By combining the submm LABOCA and MIR MSX 
data one can distinguish between candidate starless and star-forming clumps.
Of the twelve clumps in G304.74, four are likely to be associated with newly
born stars. In addition to the three IRAS (and MSX) sources, the clump SMM 6
is associated with two MSX 8 $\mu$m point sources. The remaining eight MIR 
dark clumps are either starless or contain low-luminosity (low-mass) YSOs 
falling under the detection limit of the MSX (cf. \cite{parsons2009}).
For instance, the Spitzer/GLIMPSE survey with better sensitivity and 
resolution compared to that of MSX, revealed that some IRDC clumps 
previously thought to be starless do contain MIR sources 
(e.g., \cite{chambers2009}). Moreover, Chambers et al. (2009) detected H$_2$O 
masers towards some MIR dark clumps, which is a clear indication of star
formation activity in them. On the other hand, starless clumps may be either
prestellar or just unbound structures that will eventually disperse.

The extremely short lifetime of starless IRDC clumps ($10^3-10^4$ yr, i.e., 
$\sim$half of the time spent in the protostellar phase) derived
by Parsons et al. (2009) does not support the idea that all of the
eight dark clumps in G304.74 could be starless. If gravitationally
bound, these clumps with masses in the range $\sim40-200$ M$_{\sun}$ 
(Table~\ref{table:parameters}, Col.~(3)) are capable of forming high-mass 
stars (\cite{thompson2006}), and some of them could represent/contain HMSCs
(cf. \cite{chambers2009}). The peak column densities of the clumps do 
not reach the minimum column density threshold of 1 g cm$^{-2}$, i.e., 
$N({\rm H_2})\sim2.2\times10^{23}$ cm$^{-2}$, proposed by Krumholz \& McKee 
(2008) for the formation of high-mass stars. However, it is possible that the 
clumps host cores where such a high $N({\rm H_2})$ values can be reached 
(e.g., \cite{hennemann2009}).

The presence of high-luminosity ($\sim1.5-2\times10^3$ L$_{\sun}$) IRAS
sources indicates that intermediate to high-mass star formation is going on
in other parts of the cloud. In fact these luminosities
suggest intermediate-mass stars as high-mass protostellar objects (HMPOs)
are often found in the range $\sim10^{3.5}-10^{5.5}$ L$_{\sun}$ 
(e.g., \cite{sridharan2002}; \cite{fazal2008}; \cite{grave2009}). 
On the other hand, the envelope masses of these sources ($\sim60-90$ 
M$_{\sun}$, see Col.~(2) of Table~\ref{table:sed}) are 
sufficiently large for high-mass star formation (\cite{beutherstein2007}). 
For comparison, majority of the low-mass protostars in nearby 
($d\lesssim250$ pc) molecular clouds have luminosites 
$L_{\rm bol}\leq{\rm a\, few}\times {\rm L}_{\sun}$, 
the highest observed values being $\sim70$ L$_{\sun}$ in only a few cases 
(e.g., \cite{evans2009}). IRAS 13037-6112 and IRAS 13039-6108 have 
similar SED parameters (see Fig.~\ref{figure:sed} and Table~\ref{table:sed}), 
suggesting that they probably represent the same evolutionary stage. 
Based on its IRAS $[25-12]$ and $[60-12]$ colours (0.61 and 1.55, 
respectively), IRAS 13037-6112 belongs to the so-called \textit{High} sources,
and could potentially be associated with ultra-compact (UC) 
H{\footnotesize II} region (e.g., \cite{wood1989}; \cite{palla1991}; 
\cite{molinari1996}). However, the $M/L$ ratio of IRAS 13037-6112 (0.04) is 
more typical of sources younger than UC H{\footnotesize II} regions 
(\cite{sridharan2002}). IRAS 13039-6108 has a $[25-12]$ colour index of 0.27,
settling it to the so-called \textit{Low} sources. Moreover, 
Fontani et al. (2005) found that the C$^{17}$O and CS linewidths in 
IRAS 13039-6108 are significantly smaller than those typically observed in 
massive clumps associated with UC H{\footnotesize II} regions 
(e.g., \cite{cesaroni1991}; \cite{hofner2000}). 
This further supports the idea that both IRAS 13037-6112 
and IRAS 13039-6108 represent the same evolutionary stage 
(earlier than UC H{\footnotesize II}). Using the assumption of an isothermal 
sphere with $T_{\rm kin}=22$ K and a density profile of the form 
$n(r)\propto r^{-1.6}$, which is typical of high-mass star-forming clumps
(e.g., \cite{beuther2002a}), the C$^{17}$O$(2-1)$ linewidth of 
0.93 km s$^{-1}$ observed by Fontani et al. (2005) implies a virial mass
of $\sim99$ M$_{\sun}$ for IRAS 13039-6108 
($M_{\rm cont}/M_{\rm vir}\simeq0.8$, $M_{\rm SED}/M_{\rm vir}\simeq0.9$, 
i.e., the virial parameter defined by Bertoldi \& McKee (1992) is 
$\alpha_{\rm vir}=M_{\rm vir}/M\simeq1.1-1.3$) 
(see, e.g., Eqs.~(1) and (2) of \cite{chen2008}). Thus the clump is 
near virial equilibrium. IRAS 13042-6105 in the northern part of the cloud is 
probably in an earlier stage of evolution than the other two IRAS sources 
because it is not as bright at FIR wavelengths (see Table~\ref{table:IRAS}).

\subsection{Clump mass distribution}

The mass distribution of clumps/cores is important parameter
concerning the cloud fragmentation mechanism.
Our sample of clumps is, however, so small (12 in total, 8 MIR dark) that it
is not reasonable to study their mass distribution directly. 
Therefore, we only compared it with the mass distributions derived for other,
larger IRDC clump samples, using the results of Sridharan et al. (2005), 
RJS06, Vasyunina et al. (2009), and RBG09. 

\begin{figure}[!h]
\resizebox{\hsize}{!}{\includegraphics{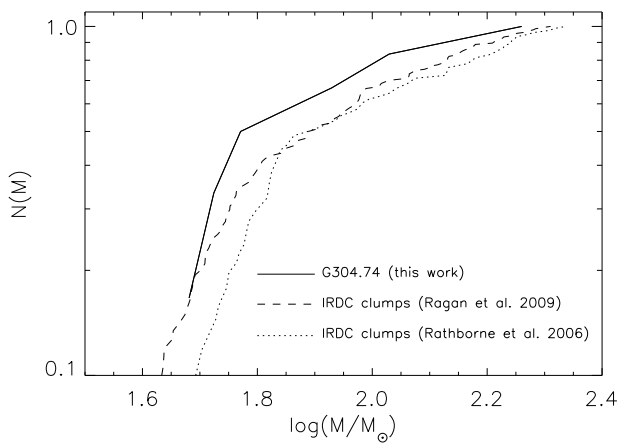}}
\caption{Normalised cumulative mass functions, $\mathcal{N}(M)$,
for the MIR dark clumps in the IRDC G304.74 (solid line) and in 
several other IRDCs studied by RJS06 (dotted line), and by RBG09 
(dashed line). Clump masses from RJS06 and RBG09 were scaled down 
to match our assumptions about $T_{\rm d}$ and $\kappa_{\lambda}$ 
(see main text)}.
\label{figure:CMF}
\end{figure}

Figure~\ref{figure:CMF} presents the observed cumulative mass functions,
which include clumps of mass less than $M$, i.e., 
$\mathcal{N}(M)=N(m<M)/N_{\rm tot}$, for G304.74 and for a sample of cold 
clumps from RJS06 and RBG09.
We note that the IRDC clump mass functions in RJS06 and in our work 
are constructed by removing the MIR bright clumps from the samples 
(in the case of G304.74, this means the three IRAS sources and SMM 6). 
From the sample of Sridharan et al. (2005), we excluded the high temperature 
clump No.~9, because its temperature (32.7 K) was much higher than the rest 
of the sample. From the the sample of RBG09, 
we removed the clumps associated with YSOs, and the clumps possibly 
contaminated by foreground (or background) stars. Thus, 
the mass functions include only those clumps that initially have all their 
mass available for star formation. 

The previous studies taken into this comparison used slightly
different assumptions about the dust temperature and opacity.
We scaled all the clump masses included in the cumulative mass functions
to correspond to a uniform dust temperature of $T_{\rm d}=15$ K, and an opacity
that is consistent with our $\kappa_{870}=0.17$ m$^2$ kg$^{-1}$.
RBG09 derived clump masses from the total hydrogen column
densities, $N_{\rm H}$, as estimated from the 8 $\mu$m optical thicknesses, 
$\tau_8$. For the dust extinction cross-section per H nucleus at 8 $\mu$m, 
$\sigma_8$, they used the value $2.3\times10^{-23}$ cm$^2$, based on the 
Weingartner \& Draine (2001) dust model. According to the model we have used 
(\cite{ossenkopf1994}), the corresponding number is $3.5\times10^{-23}$ cm$^2$
per H nucleus (see Sect. 5.3). In their mass formula, RBG09 used a factor of 
1.16 as the ratio of the total gas mass (including He) and the hydrogen
mass. In our calculations this ratio has been 1.4 (Sect. 5.3). These
differences have been accounted for by scaling the masses from RBG09 by 
0.8 in the comparison with our results.

\begin{table}
\caption{K-S test results between different IRDC clump mass distributions.}
\begin{minipage}{1\columnwidth}
\centering
\renewcommand{\footnoterule}{}
\label{table:KS}
\begin{tabular}{c c c c}
\hline\hline 
 & Number of &  & \\  
Study & cold clumps & $D_{\max}$ & Prob.\\
\hline
Ragan et al. (2009, RBG09)\footnote{The probability that the RBG09 and
RJS06 clump mass samples represent the subsamples of the same parent 
distribution is 100\% ($D_{\max}=0.001$).} & 345 & 0.153 & 0.998\\ 
Rathborne et al. (2006, RJS06)$^a$ & 140 & 0.272 & 0.739\\
Sridharan et al. (2005) & 28 & 0.359 & 0.534\\
Vasyunina et al. (2009) & 17 & 0.449 & 0.408\\
\hline 
\end{tabular} 
\end{minipage}
\end{table}

To determine whether our clump masses and those from other studies 
are derived from the same clump mass distribution, we carried out the 
two-sample Kolmogorov-Smirnov (K-S) test. For this test, the mass scales
were matched, i.e., the comparison was done within the range of common mass
interval. The K-S test results are shown in Table~\ref{table:KS}. 
The columns of this table are: (1) the survey used in the
comparison; (2) number of clumps included; (3) the maximum vertical difference
between the cumulative mass distributions ($D_{\max}$); (4) the probability 
for the null hypothesis that the two functions are drawn form the same parent
distribution (the significance level of the K-S statistic). 

The highest number statistic is provided by the study of RBG09, 
and when compared with this study, the K-S test yields a probability of 
99.8\% that the clump mass distributions in G304.74 and the other IRDCs
are drawn from the same parent distribution. This probability drops 
significantly when smaller samples are used in the test 
(Table~\ref{table:KS}, Col.~(4)). Moreover, according to the K-S test,
there is a 100\% probability ($D_{\max}=0.001$) that the RBG09 and RJS06 clump
masses represent the subsamples of the same 
underlying parent distribution. For this test, clump masses from RBG09 were 
also multiplied by 0.8 to compare with the RJS06 masses which were based 
on the value $\kappa_{\rm 1.2mm}=0.1$ m$^2$ kg$^{-1}$. 

For the clump masses between $\sim30$ and 3000 M$_{\sun}$, RBG09
derived an IRDC clump mass spectrum with a slope of $\alpha=1.76\pm0.05$, 
which is consistent with the mass functions
derived for high-mass star-forming regions, and also resembles the mass 
function of Galactic stellar clusters (see RBG09 and references therein).
Note, however, that RBG09 merged all their clumps into a single mass function,
whereas we have used only starless clumps in deriving the cumulative mass 
functions. RJS06 found a Salpeter-like ($\alpha=2.35$; \cite{salpeter1955})
mass function for their cold IRDC clumps 
($\alpha=2.1\pm0.4$ at $M\gtrsim100$ M$_{\sun}$).
The IRDC clump mass spectra derived by RJS06 and RBG09 are 
comparable to those predicted by turbulent 
fragmentation models (see Sect.~6.7 and references therein).

\subsection{Clump spatial distribution}

In addition to the clump mass distribution, it is also useful to 
determine how clumps are spatially distributed within the cloud in order 
to unveil the presence of a possible preferred length-scale of 
fragmentation (e.g., \cite{munoz2007}).
For this purpose, we determined the clump-separation distributions and 
the number distributions of the projected separation distance between 
nearest neighbours\footnote{If the number of clumps in the cloud is 
$N_{\rm cl}$, then the number of clump-separations is 
$N_{\rm cl}(N_{\rm cl}-1)/2$, whereas the number of distances between an 
individual clump and its nearest neighbour is equal to $N_{\rm cl}$.} 
in G304.74, and in nine other IRDCs for comparison.
We chose those IRDCs from the sample of RJS06 which contain 
the largest number of clumps, i.e., MSXDC G023.60+00.00, 
G024.33+00.11, G028.37+00.07, G028.53-00.25, G031.97+00.07, G033.69-00.01, 
G034.43+00.24, and G035.39-00.33. Moreover, we determined the spatial 
distribution of YSOs in the IRDC MSXDC G048.65-00.29 studied by van der Wiel
\& Shipman (2008). For these analyses, we selected only those sources 
that are clearly associated (in the plane of the sky) with their parental dark
cloud (e.g., clumps that lie within the dark filaments). 
Thus, we excluded the millimetre clumps MM 1, 3, and 5 from G023.60+00.00,
MM 2, 5, and 7 from G024.33+00.11, MM 3, 5, 7, 8, 12, 13, 18 from 
G028.37+00.07, MM 1 and 2 from G035.39-00.33, and YSOs S4, 10, 17, 18, 19,
and 20 from G048.65-00.29. In addition to the observed spatial 
distributions, we also determined the distributions expected from random 
positions of the same number of objects as the observed samples have. The 
areas over which the objects were randomly distributed were chosen so that 
they approximate the observed dark cloud areas; the IRDC areas were estimated 
by rectangles which just cover the observed dark clouds. When needed, 
these rectangles were rotated with respect to the $(\alpha,\delta)$-coordinate
system. The random distributions were generated a hundred times per
cloud and the resulting averaged histograms were used in comparisons
with observed spatial distributions.

\begin{figure}[!h]
\resizebox{\hsize}{!}{\includegraphics{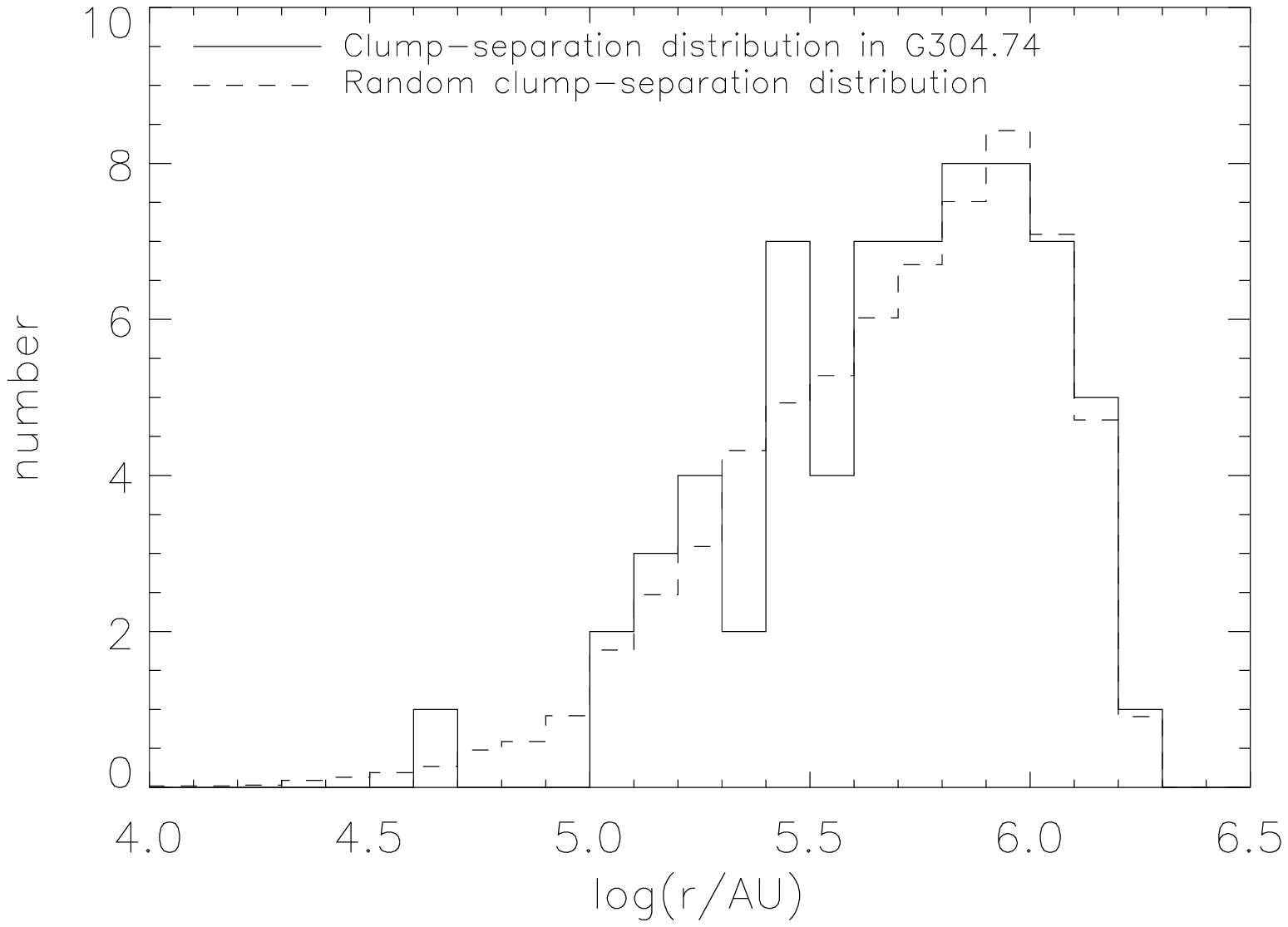}}
\resizebox{\hsize}{!}{\includegraphics{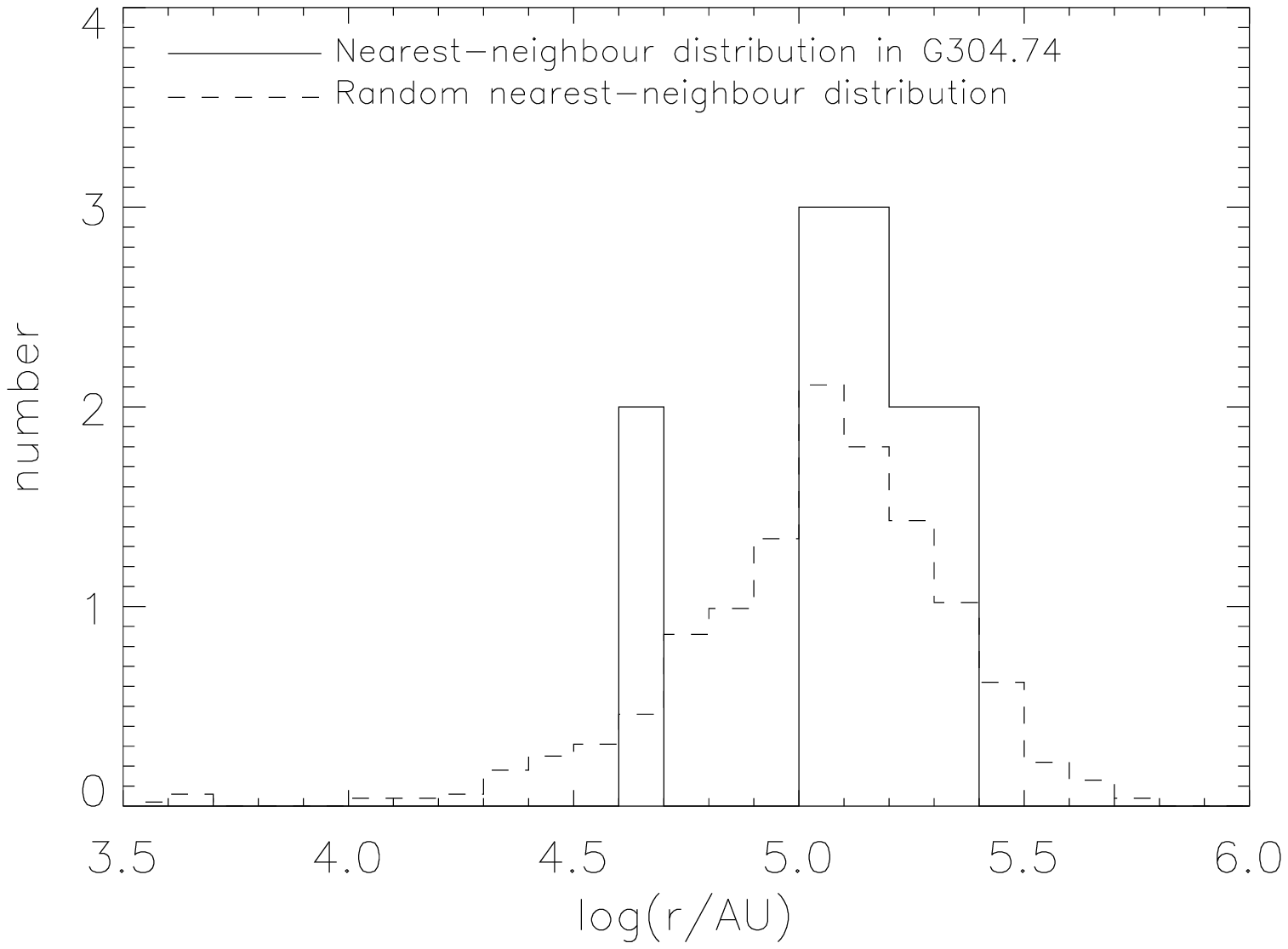}}
\caption{{\bf Top:} Observed clump-separation distribution (solid line) 
compared with the expected distribution for random distribution of the same
number of clumps as the observed sample over rectangular area which
approximate the area of the IRDC (dashed line).
{\bf Bottom:} Observed nearest-neighbour distribution (solid line)
compared with the expected distribution for random distribution (dashed line).}
\label{figure:dist}
\end{figure}

Figure~\ref{figure:dist} (top) shows the observed clump-separation distribution
in G304.74, and the distribution expected for the same number of
randomly positioned clumps over minimum rectangular area which encloses the 
dark cloud ($\sim29.8$ arcmin$^2$). The mean and its standard 
deviation, and median of the clump separations in G304.74 are 
$\log(\langle r \rangle_{\rm obs}/{\rm AU})=5.690\pm0.041$ 
($4.90^{+0.48}_{-0.44}\times10^5$ AU) and 
$\log(\tilde{r}_{\rm obs}/{\rm AU})=5.759$ ($5.74\times10^5$ AU), 
respectively. These values are similar to those of randomly 
positioned clumps, for which the mean and median are 
$\log(\langle r \rangle_{\rm ran}/{\rm AU})=5.674\pm0.061$ and 
$\log(\tilde{r}_{\rm ran}/{\rm AU})=5.738\pm0.084$ (see 
Table~\ref{table:separations}). The latter two values and their $\pm$-errors 
quoted represent the average values and their standard deviations derived from
the 100 generated random distributions. According to the K-S test, the 
probability that the observed distribution and the generated random 
distribution represent the same underlying distribution is 100\%.
Statistics of the clump-separation distributions in other IRDCs studied by 
RJS06 and van der Wiel \& Shipman (2008) are listed in 
Table~\ref{table:separations}. The columns of this table are the following: 
(1) IRDC name; (2) number of clumps used in the analysis (see above); 
(3) distance; (4) area used to create the random distributions 
(see above); (5) and (6) mean and median of the observed clump-separation 
distribution ($\langle r \rangle_{\rm obs}$ and 
$\tilde{r}_{\rm obs}$); (7) and (8) mean and median of the corresponding 
random distribution ($\langle r \rangle_{\rm ran}$ and 
$\tilde{r}_{\rm ran}$); (9) and (10) ratios between the observed and random 
mean and median separations (quoted errors are propagated from the standard 
deviations of $\langle r \rangle$ and $\tilde{r}$); (11) probability given by 
the K-S test that the observed and random distributions are drawn from the same
underlying distribution. The observed clump separations are mostly similar to 
those expected from random distributions. This is evident from the ratios 
$\langle r \rangle_{\rm obs}/\langle r \rangle_{\rm ran}$ 
and $\tilde{r}_{\rm obs}/\tilde{r}_{\rm ran}$ which are close to unity, and 
from the K-S probabilities which are high ($\sim71-100\%$) except for three 
cases (G035.39, G028.37, G024.33; see Cols.~(9)-(11) of 
Table~\ref{table:separations}).

Figure~\ref{figure:dist} (bottom) compares the observed 
nearest-neighbour distribution in G304.74 with the distribution for
randomly positioned clumps. The mean and median of the 
nearest-neighbour distribution in G304.74 are 
$\log(\langle r \rangle_{\rm obs}/{\rm AU})=5.083\pm0.058$ 
($1.21^{+0.17}_{-0.15}\times10^5$ AU) and 
$\log(\tilde{r}_{\rm obs}/{\rm AU})=5.136$ ($1.37\times10^5$ AU), 
respectively. Again, these values are comparable to those of randomly 
positioned clumps, for which the mean and median are 
$\log(\langle r \rangle_{\rm ran}/{\rm AU})=5.030\pm0.111$ and 
$\log(\tilde{r}_{\rm ran}/{\rm AU})=5.047\pm0.110$, respectively 
(see Table~\ref{table:nearest}). 
According to the K-S test, there is about 90\% probability that the 
observed and random nearest-neighbour distributions are samples of the same 
underlying distribution. We note that the minimum observable separation 
corresponds to the beam size, i.e, $18\farcs6$ or $\sim4.46\times10^4$ AU 
($\log(r/{\rm AU})=4.649$) at 2.4 kpc. Statistics of the nearest-neighbour 
distributions in other IRDCs are given in Table~\ref{table:nearest}. 
Table~\ref{table:nearest} have the same meaning as in 
Table~\ref{table:separations}, except now for nearest neighbour separations. 
The observed nearest-neighbour distances are similar to those expected from 
random distributions. This is evident from 
the ratios $\langle r \rangle_{\rm obs}/\langle r \rangle_{\rm ran}$ and 
$\tilde{r}_{\rm obs}/\tilde{r}_{\rm ran}$ which are (within the erros) about 1,
and by the high K-S probabilities ($\sim59-100\%$) in every other case except 
G031.97, where this probability is still $\sim37\%$ (see Cols.~(6)-(8) of 
Table~\ref{table:nearest}).

In summary, the average projected separations between clumps in the
studied IRDCs range from about $2.6\times10^5$ AU to $1.2\times10^6$ AU (i.e., 
the minimum and maximum lie within a factor of five), and the average
projected distances between the nearest neighbours range from $6.0\times10^4$
AU to $3.5\times10^5$ AU (i.e., the changes are within a factor of six). For
most clouds, the distributions of projected separations and distances
between the nearest neighbours can be mimicked by clumps placed
randomly into the same projected area as occupied by the cloud.
Assuming that the vectors connecting clump pairs are randomly oriented, 
the average projection factor is 
$\langle \sin \theta_{ij} \rangle=\pi/4$, where $\theta_{ij}$ is the 
angle between the line of sight and the vector pointing from clump $i$ to 
clump $j$. Correcting for this projection effect, the grand averages of the 
separations and distances between the nearest neighbours are about 
$6.5\times10^5$ AU (3.1 pc) and $2.2\times10^5$ AU (1.1 pc), respectively.

\subsection{Fragmentation of IRDCs}

One plausible scenario for the origin of filamentary clouds  
is that they are formed in shocks occurring in converging flows driven 
by large-scale turbulence (e.g., \cite{klessen2000}; \cite{padoan2001}). 
The chaotic process can give rise to 
randomly positioned density peaks within filaments, and these can become 
centres of gravitational collapse. On the other hand, supposing that 
compression leads to an equilibrium structure, a filament can 
fragment through the Jeans instability. 

The critical wavelength, $\lambda_{\rm c}$, of perturbations leading
to gravitational instability depends on both the gas
kinetic tempeture, $T_{\rm kin}$, and the density, $\rho$:
$\lambda_{\rm c} \sim c_{\rm s}/\sqrt{G\rho}$, or, in terms of the
surface density, $\Sigma$: $\lambda_{\rm c} \sim c_{\rm s}^2/{G\Sigma}$, 
where $c_{\rm s}$ is the sound speed, and $G$ is the
gravitational constant (e.g., \cite{larson1985}; \cite{hartmann2002}).

The determination of the 'Jeans length' is not quite straightforward
in a study based on dust emission because the cloud mass and therefore 
also the average density and surface density depend on the assumed dust
temperature, $T_{\rm d}$. Furthermore, in the case of G304.74, there is 
no independent estimate of $T_{\rm kin}$, but it is assumed to be equal 
to $T_{\rm d}$. 

The assumption $T_{\rm d}=15$ K yields a total mass of 
$\sim1000$ M$_{\sun}$ and an average surface density of 0.05 g\,cm$^{-2}$ for
G304.74 (within the LABOCA contour 0.1 Jy beam$^{-1}$). In these
circumstances, the critical wavelength in an isothermal equilibrium filament is
$\lambda_{\rm c}=0.19$ pc or 40\,000 AU (\cite{hartmann2002}; 
\cite{larson1985}), and the corresponding mass is $M_{\rm c}\sim5$ M$_{\sun}$.
The comparison between dust emission at 870 $\mu$m, 8 $\mu$m absorption,
and the visual extinction from 2MASS suggests an elevated temperature
in the southern part of the cloud (Table~\ref{table:extinction}, Col.~(4)).
By assuming $T_{\rm d}=30$ K, one
would obtain a total cloud mass of $\sim400$ M$_{\sun}$, and an average
surface density of $\Sigma = 0.02$ g\,cm$^{-2}$. These values of
temperature and surface density imply $\lambda_{\rm c} = 1.1$ pc or
$2.2\times10^5$ AU, and $M_{\rm c} \sim 50$ M$_{\sun}$. The critical 
wavelength, $\lambda_{\rm c}$, is not expected to determine a uniform length 
scale of fragmentation. According to the analysis of Stod{\'o}lkiewicz (1963;
see also \cite{curry2000} and references therein), the fastest growing
perturbations have a length scale of roughly twice $\lambda_{\rm c}$.
Nevertheless, the fragmentation of a homogenous cloud is likely to result 
in a preferred length scale and a quasi-periodic structure, as opposed to 
a random distribution of clumps. 
 
The projected distances between the nearest neighbours in G304.74 lie
in the range $4\times10^4 - 2.4\times10^5$ AU. They are comparable to the
characterics length scales indicated above. However, the clump masses
in the cool northern part ($\sim40-90$ M$_{\sun}$) are about ten times
larger than Jeans masses at 15 K and the surface density implied by
this temperature. Either the Jeans instability has occurred when the
cloud has been warmer, i.e. $\sim30$ K, and the clumps have cooled
during the contraction, or the dense filament is a result of strong
compression by external forces. We note that the filament is thinner
at the northeastern end (cross-sectional diameter $\phi \sim 0.45$ pc)
than in the southwest ($\phi \sim 1.4$ pc). Moreover, the filament can
have accumulated mass through gravitational inflow from the
surrounding cloud (\cite{heitsch2009}).

The present average mass line density in G304.74, 
$\sim 100$ M$_{\sun}$ pc$^{-1}$ exceeds the critical value for a non-magnetic, 
self-gravitating isothermal cylinder in equilibrium 
($\sim25$ M$_{\sun}$ pc$^{-1}$ at 15 K, the critical line density directly 
proportional to $T_{\rm kin}$; 
\cite{ostriker1964}; \cite{curry2000}). For comparison, in several
filamentary IRDCs studied by Rathborne et al. (2006; G025.04, G028.53,
G028.53, G031.97, G033.69, G034.43, and G035.39), the line densities
are in the range $\sim70 - 800$ M$_{\sun}$ pc$^{-1}$. Supercritical line 
densities seem to be a common feature in filamentary IRDCs, and they are 
likely to fragment into smaller cores. Our observations cannot resolve the
possible fragments in G304.74, but the process is manisfest in
presence of (intermediate- to high-mass) protostars, i.e., the two IRAS 
sources. On the other hand, the recent interferometric studies of
G28.34 (\cite{zhang2009}) and IRDC 19175 (\cite{beuther2009})
have provided direct evidence for sub-fragmentation of IRDC clumps.
The two studies offer, however, contradictory views of the nature of
the underlying instability (turbulent fragmentation vs. Jeans instability).

\begin{table*}
\caption{Statistics of the clump-separation distributions in IRDCs.}
\begin{minipage}{2\columnwidth}
\centering
\renewcommand{\footnoterule}{}
\label{table:separations}
\begin{tabular}{c c c c c c c c c c c}
\hline\hline 
       &     &      & & \multicolumn{2}{c}{\textbf{Observed distribution}} & \multicolumn{2}{c}{\textbf{Random distribution}}\\ 
Name MSXDC & $N_{\rm cl}$ & $d$ & Area & $\langle r \rangle_{\rm obs}$ & $\tilde{r}_{\rm obs}$ & $\langle r \rangle_{\rm ran}$ & $\tilde{r}_{\rm ran}$ & $\langle r \rangle_{\rm obs}/\langle r \rangle_{\rm ran}$ & $\tilde{r}_{\rm obs}/\tilde{r}_{\rm ran}$ & Prob.\\ 
  & & [kpc] & [$\Box \arcmin$] & [$\log$ AU] & [$\log$ AU] & [$\log$ AU] & [$\log$ AU] & & & \\  
\hline
G304.74+01.32 & 12 & 2.4 & 29.8 & $5.690\pm0.041$ & 5.759 & $5.674\pm0.061$ & $5.738\pm0.084$ & $1.04\pm0.18$ & $1.05\pm0.20$ & 1.000\\
G048.65-00.29 & 14\footnote{Number of YSOs.} & 2.5 & 13.5 & $5.482\pm0.035$ & 5.524 & $5.449\pm0.059$ & $5.497\pm0.071$ & $1.08\pm0.17$ & $1.06\pm0.17$ & 0.709\\
G035.39-00.33 & 7 & 2.9 & 17.2 & $5.422\pm0.058$ & 5.493 & $5.583\pm0.095$ & $5.630\pm0.117$ & $0.69\pm0.18$ & $0.73\pm0.20$ & 0.156\\
G034.43+00.24 & 9 & 3.7 & 12.6 & $5.608\pm0.051$ & 5.672 & $5.618\pm0.075$ & $5.670\pm0.096$ & $0.98\pm0.21$ & $1.00\pm0.22$ & 1.000\\
G033.69-00.01 & 11 & 7.1 & 28.7 & $6.089\pm0.044$ & 6.152 & $6.076\pm0.062$ & $6.123\pm0.083$ & $1.03\pm0.18$ & $1.05\pm0.20$ & 0.982\\
G031.97+00.07 & 9 & 6.9 & 18.1 & $5.861\pm0.051$ & 5.870 & $5.943\pm0.073$ & $5.995\pm0.088$ & $0.83\pm0.17$ & $0.75\pm0.15$ & 0.738\\
G028.53-00.25 & 10 & 5.7 & 17.6 & $5.733\pm0.042$ & 5.797 & $5.822\pm0.060$ & $5.879\pm0.070$ & $0.81\pm0.14$ & $0.83\pm0.13$ & 0.753\\
G028.37+00.07 & 11 & 5.0 & 51.5 & $5.903\pm0.033$ & 5.934 & $5.977\pm0.055$ & $6.044\pm0.065$ & $0.84\pm0.13$ & $0.78\pm0.12$ & 0.324\\
G024.33+00.11 & 8 & 3.8 & 27.5 & $5.772\pm0.061$ & 5.926 & $5.727\pm0.073$ & $5.789\pm0.085$ & $1.11\pm0.24$ & $1.37\pm0.27$ & 0.152\\
G023.60+00.00 & 6 & 3.9 & 7.9 & $5.506\pm0.048$ & 5.480 & $5.491\pm0.099$ & $5.542\pm0.114$ & $1.04\pm0.26$ & $0.87\pm0.23$ & 0.822\\
\hline 
\end{tabular} 
\end{minipage}
\end{table*}

\begin{table*}
\caption{Statistics of the number distributions of the projected separation 
distance between nearest neighbours in IRDCs.}
\begin{minipage}{2\columnwidth}
\centering
\renewcommand{\footnoterule}{}
\label{table:nearest}
\begin{tabular}{c c c c c c c c}
\hline\hline 
       & \multicolumn{2}{c}{\textbf{Observed distribution}} & \multicolumn{2}{c}{\textbf{Random distribution}}\\ 
Name MSXDC & $\langle r \rangle_{\rm obs}$ & $\tilde{r}_{\rm obs}$ & $\langle r \rangle_{\rm ran}$ & $\tilde{r}_{\rm ran}$ & $\langle r \rangle_{\rm obs}/\langle r \rangle_{\rm ran}$ & $\tilde{r}_{\rm obs}/\tilde{r}_{\rm ran}$ & Prob.\\ 
  & [$\log$ AU] & [$\log$ AU] & [$\log$ AU] & [$\log$ AU]& & & \\  
\hline
G304.74+01.32 & $5.083\pm0.058$ & 5.136 & $5.030\pm0.111$ & $5.047\pm0.110$ & $1.13\pm0.33$ & $1.23\pm0.31$ & 0.901\\
G048.65-00.29 & $4.786\pm0.046$ & 4.741 & $4.854\pm0.108$ & $4.875\pm0.108$ & $0.85\pm0.23$ & $0.73\pm0.18$ & 0.860\\
G035.39-00.33 & $5.123\pm0.093$ & 4.999 & $5.145\pm0.125$ & $5.137\pm0.161$ & $0.95\pm0.34$ & $0.73\pm0.27$ & 0.823\\
G034.43+00.24 & $5.117\pm0.036$ & 5.081 & $5.104\pm0.128$ & $5.123\pm0.143$ & $1.03\pm0.31$ & $0.91\pm0.30$ & 0.966\\
G033.69-00.01 & $5.539\pm0.019$ & 5.528 & $5.512\pm0.100$ & $5.523\pm0.119$ & $1.06\pm0.25$ & $1.01\pm0.28$ & 0.594\\
G031.97+00.07 & $5.447\pm0.066$ & 5.354 & $5.450\pm0.128$ & $5.469\pm0.148$ & $0.99\pm0.33$ & $0.77\pm0.26$ & 0.366\\
G028.53-00.25 & $5.281\pm0.077$ & 5.356 & $5.334\pm0.118$ & $5.360\pm0.123$ & $0.88\pm0.29$ & $0.99\pm0.28$ & 0.843\\
G028.37+00.07 & $5.520\pm0.073$ & 5.597 & $5.490\pm0.096$ & $5.501\pm0.113$ & $1.07\pm0.30$ & $1.25\pm0.32$ & 1.000\\
G024.33+00.11 & $5.236\pm0.040$ & 5.230 & $5.308\pm0.117$ & $5.314\pm0.148$ & $0.85\pm0.24$ & $0.82\pm0.28$ & 0.792\\
G023.60+00.00 & $5.280\pm0.057$ & 5.341 & $5.142\pm0.148$ & $5.153\pm0.172$ & $1.37\pm0.50$ & $1.54\pm0.61$ & 0.780\\
\hline 
\end{tabular} 
\end{minipage}
\end{table*}

\subsection{The origin of IRDCs and substructures within them}

IRDCs are the densest parts of molecular cloud complexes. Several
observational facts and modelling results suggest that turbulence has
an important role in the formation of IRDCs and their fragmentation
into clumps. 1) On large scales, molecular clouds have highly
supersonic linewidths. In general, these are considered to imply
turbulent motions (e.g., \cite{elmegreen2004}; \cite{mckee2007}). 
However, large linewidths in dense clouds can also be
explained by collapsing motions towards local gravitational centres
(\cite{heitsch2009}). 2) The filamentary shapes of IRDCs, and molecular
clouds in general, are consistent with cloud morphologies predicted by
numerical models of supersonic turbulence driven on large scales
(e.g., \cite{klessen2000a}, 2001; \cite{jappsen2005}). In this
model, a dense filament can form where converging flows meet. Also
the mass surface density distribution observed in IRDCs have been
found to correspond to expectations from numerical simulations of
turbulent clouds (\cite{butler2009}). 3) The high-mass end of the mass
distiribution of a large sample of IRDCs studied by Marshall et al.
(2009) follows a power-law (${\rm d}N/{\rm d}M \propto M^{-1.75}$) which can be
reproduced by density fluctuations induced by turbulence. 4) The mass
spectrum of the high-mass clumps within the extensive IRDC sample of
RJS06 can be fitted with a power-law (${\rm d}N/{\rm d}M \propto M^{-2.1}$). 
This power-law agrees with fragmentation due to supersonic turbulence in
self-gravitating clouds (\cite{klessen2001}). 5) Finally, the spatial
distributions of clumps within IRDCs (cf. Sect. 6.5) show no clear
deviation from a random distribution. This is what can be expected if
fragmentation is driven by a stochastic process.

The filamentary stuctures of IRDCs and the fact that star formation in
them takes place in clusters suggest that turbulence is driven on
large scales (e.g., \cite{klessen2001}). The most likely driving agent for
large-scale turbulence is provided by supernova explosions, with an
important contribution of density fluctuations caused by older
remnants (e.g., \cite{korpi1998}; \cite{joung2009} and references
therein). The Galactic distribution of IRDCs peak in the 5-kpc
molecular ring (\cite{simon2006a}; \cite{marshall2009}), where most
of the Galactic supernova remnants are found (e.g., \cite{jackson2008}).

The role of turbulence in the fragmention of IRDC clumps into dense
cores is less evident. The observed spectral linewidths in clumps are
broader than expected from thermal broadening (e.g., \cite{ragan2006};
\cite{sakai2008}; \cite{gibson2009}), but this does not inevitably
imply that the gas is turbulent (\cite{heitsch2009}). As discussed in
Sect. 6.6, high-resolution studies have revealed some regular
stuctures within clumps (\cite{zhang2009}; \cite{beuther2009})
which point towards gravitational or fluid dynamical instabilities.
In any case, self-gravity of shock-compressed clumps or filaments 
is inherent also in the models of turbulent fragmentation. 

\section{Summary and conclusions}

We have mapped IRDC G304.74 in the 870 $\mu$m dust continuum emission
with the APEX telescope. The submm dust continuum observations have been used
together with 1.2 mm data from Beltr{\'a}n et al. (2006), and archival MSX
and IRAS infrared data to derive the physical characteristics of the 
clumps within the cloud. Besides the dust continuum we used dust 
extinction data from MSX and 2MASS to derive the H$_2$  column densities, and 
the mass distribution in the cloud. The results obtained via different methods 
are in reasonable agreement with each other taking into account the uncertain 
nature of some of the dust properties and the relation between the H$_2$ 
column density and extinction. However, the agreement can be improved by 
assuming an elevated temperature in four clumps near the southwestern end of 
the cloud, and the possibility of a temperature gradient from about 15 K
in the north and centre to about 20-30 K in the south cannot be ruled out.

The filamentary cloud G304.74 contains 12 submm clumps. 
Star formation has already started in the cloud as three of the 
clumps are associated with both MSX and IRAS point sources. 
The SEDs of the two IRAS sources indicate bolometric luminosities 
in the range $\sim1.5-2\times10^3$ L$_{\sun}$. These are likely to be 
intermediate- or high-mass protostars.
In addition, one of the clumps (SMM 6) is associatied with two MSX 8 $\mu$m 
point sources. The remaining eight submm clumps are MIR dark. 
The masses of these clumps ($\sim40-200$ M$_{\sun}$)
are sufficiently large to enable high-mass star formation, or 
alternatively, they can represent the cold precursors of proto-clusters. 
Thus, some of the candidate starless clumps in G304.74 could represent/harbour 
the so-called high-mass starless cores (HMSCs, e.g., \cite{beuther2007}).
Further studies of these high-mass starless clump/core
candidates are important in order to constrain the initial conditions of 
high-mass star and star cluster formation. 

The clump masses in G304.74 were compared with the clump mass spectra
from more extensive surveys of IRDCs. We found that IRDC clump 
masses from the present work, and those from Rathborne et al. (2006) and Ragan
et al. (2009) probably represent subsamples of the same parent distribution. 
Also, average distances between a clump and its nearest neighbour in
different IRDCs are comparable (within a factor of $\sim6$), 
suggesting that the fragmentation length-scale does not vary much from cloud
to cloud. Moreover, in most IRDCs, clumps seem to be randomly 
distributed within the cloud area.
These characteristics, and the fact the star formation in IRDCs predominantly
occurs in the cluster mode, agree with models where fragmentation is
driven by large-scale turbulence. It is not clear, yet, how effectual 
turbulence is for the fragmentation of IRDC clumps into dense cores. High 
spatial resolution studies, like the ones presented in Zhang et al. (2009) and 
Beuther \& Henning (2009) have recently started to throw light on scales where 
gravity is likely to dominate.

\begin{acknowledgements}

The authors thank the referee for comments and suggestions which significantly 
improved the paper.
We would like to thank the APEX staff in Chile for performing the LABOCA 
observations. We acknowledge M. Hennemann for providing the SED fitting tool 
originally written by J. Steinacker. Furthermore, M.~T. Beltr{\'a}n, 
P. Bergman, M.~J. Butler, J. Kainulainen, H. Linz, and K. Mattila are
thanked for useful discussions and suggestions. The authors acknowledge 
support from the Academy of Finland through grants 117206 and 132291.
This research made use of data products from the Midcourse Space Experiment. 
Processing of the MSX data was funded by the Ballistic Missile Defense 
Organization with additional support from the NASA Office of Space Science. 
In addition, this publication makes use of data products from the Two Micron 
All Sky Survey, which is a joint project of the University of Massachusetts 
and the Infrared Processing and Analysis Center/California Institute of 
Technology, funded by the National Aeronautics and Space Administration and 
the National Science Foundation. This work has made use of the 
NASA/IPAC Infrared Science Archive, which is operated by the Jet Propulsion 
Laboratory/California Institute of Technology, under contract with NASA, 
the NASA Astrophysics Data System, and the VizieR Catalogue access tool 
(CDS, Strasbourg, France).

\end{acknowledgements}

\end{document}